\documentstyle{ar}
\input epsf
\begin{document}
%% AUTHOR,  START TYPING FROM HERE 
\title{ THE EVOLUTION OF ROTATING STARS}  
\markboth{André MAEDER and Georges MEYNET}
{THE EVOLUTION OF ROTATING STARS}
\author{André MAEDER and Georges MEYNET\affiliation{Geneva Observatory,
CH--1290 Sauverny, Switzerland\\ email: andre.maeder@obs.unige.ch\\
email: georges.meynet@obs.unige.ch}}


\begin{abstract}
First, we review the main physical effects to be considered 
in the building of evolutionary models of rotating stars on the
Upper Main-Sequence (MS). The
internal rotation law evolves as a result of contraction and expansion, meridional 
circulation, diffusion processes and mass loss. In turn,  
differential rotation and mixing exert a feedback
on circulation and diffusion, so that a consistent treatment
is  necessary. 

We review recent results on the evolution of internal
rotation and the  surface rotational velocities 
for stars on the Upper MS, for red giants, supergiants and W--R stars.
A fast rotation is enhancing the mass loss by stellar winds
and reciprocally high mass loss is removing a lot of angular 
momentum.
The problem of the ``break--up'' or $\Omega$--limit
is critically examined in connection with the origin of
Be and LBV stars.
The effects of rotation on the tracks in the HR diagram, 
the lifetimes, the isochrones, the blue to red supergiant
ratios, the formation of W--R stars, 
the chemical abundances in massive stars as well as 
in red giants and AGB stars, are reviewed in relation to
recent observations for stars in the Galaxy and  Magellanic 
Clouds. The effects of rotation on the final stages and on the 
chemical yields are examined, as well as the constraints placed
by the periods of pulsars.
On the whole, this review points out that stellar evolution 
is not only a function of mass M and metallicity Z, but of
angular velocity $\Omega$ as well.

\end{abstract}

\maketitle

\section{INTRODUCTION}

Stellar rotation is an example of an astronomical domain which has been studied for
several centuries and where the
developments are rather slow.  
A  short historical review since the discovery of the solar 
rotation by Galileo Galilei is given by Tassoul (1978). 
Few of the early works apply to real stars, since in general
gaseous configurations were  not considered and  no account was
given to
radiative energy transport. The equations of rotating stars in radiative
equilibrium were first considered by Milne (1923), von Zeipel (1924)
and Eddington (1925); see also Tassoul (1990) for a 
more recent history. From the early days of stellar evolution,
the studies of
rotation and evolution have been closely associated. Soon after
the first models showing that Main--Sequence (MS) stars move further
into the giant and  supergiant region (Sandage \& Schwarzschild
1952), rotation was used as a major test for the evolution.
Oke \& Greenstein (1954) and Sandage (1955) found that the observed
rotational velocities were  consistent with the proposed evolutionary
sequence.

Stellar evolution, as other fields of Science, proceeds using
as a guideline the principle of Occam's razor, which says that
the explanation relying on the smallest number of hypotheses
is usually the one to be preferred. Thus,
due to the many well known successes of the theory of stellar 
evolution, rotation was and is still generally considered as only a 
second order effect. However, over recent years a number of 
serious discrepancies between current models and observations
have been noticed. They concern particularly the helium and 
nitrogen abundances in massive O-- and B--type stars and
in giants and supergiants, as well as the distribution of stars 
in the HR diagram at various metallicities. The
observations show that the role of rotation has
been largely overlooked. All the model outputs (tracks in the
HR diagram, lifetimes, actual masses, surface abundances,
nucleosynthetic yields, supernova precursors, etc\dots ) are
greatly influenced  by rotation, thus  it turns out that 
stellar evolution is basically
a function of mass M, metallicity Z and 
angular velocity $\Omega$.

There are a number of reviews concerning
stellar rotation, for example Strittmatter (1969), Fricke \&
Kippenhahn (1972), Tassoul (1978, 1990),  Zahn (1983, 1994) and
Pinsonneault (1997). 
Here, we focus on rotation in Upper MS stars, where the
 effects are likely the largest ones.
The consequences for
blue, yellow and red supergiants, Wolf--Rayet (W--R)
stars, as well as for red giants and Asymptotic Giant 
Branch (AGB) stars are also examined.
The rotation of low mass stars, where spin--down due to magnetic 
coupling between the wind and the central body is important,
has been treated in a recent review (Pinsonneault 1997); rotation
and magnetic activity were also reviewed by Hartmann
and Noyes (1987). The role of  rotation in  pre--Main Sequence
evolution with accretion disks has been discussed by
Bodenheimer (1995).

\section{BASIC PHYSICAL EFFECTS OF ROTATION}

\subsection{HYDROSTATIC EFFECTS}

In a rotating star, the centrifugal forces reduce the effective gravity
according to the latitude and introduce deviations from sphericity. 
The four equations of stellar structure need to be modified. The
idea of the original method devised by Kippenhahn \& Thomas (1970)
and applied in most subsequent works (Endal \& Sofia 1976; 
Pinsonneault et al 1989, 1990, 1991; Fliegner \&
Langer 1995; Heger et al 2000) is to replace the usual spherical
eulerian or lagrangian coordinates by new coordinates 
characterizing the equipotentials.
This method applies when  the effective
gravity can be derived from a potential, i.e. when the 
problem is conservative, which occurs for solid body
rotation or for constant rotation on cylinders centered on the axis
of rotation.
If so, the structural variables
$P, T, \rho, ...$ are constant on an equipotential 
$\Psi=\Phi + \frac{1}{2} \Omega^2r^2 \sin^2\vartheta$,
where $\Phi$ is the gravitational potential, $\vartheta$ 
the colatitude and $\Omega$ the angular velocity. Thus
the problem can be kept one--dimensional which is a major advantage.
However the internal rotation generally evolves towards rotation laws
which are non conservative.
In that case the above method
is not physically consistent. Unfortunately
it has been and is still used by most authors.

A particularly
interesting case  of differential rotation is that
with $\Omega$ constant on isobars (Zahn 1992).
This case is called ``shellular rotation''  and it is often 
approximated  by $\Omega = \Omega (r)$, which is valid at low 
rotation. It is supported by  the study of turbulence in the
Sun and stars (Spiegel \& Zahn 1992; Zahn 1992).
Such a law is due to the fact that
the turbulence is  very anisotropic, with
a much  stronger, geostrophic--like transport
in the horizontal direction than in the vertical one,
where stabilisation is favoured by the stable temperature gradient. 
The horizontal turbulence enforces an essentially 
constant rotation rate  
on isobars, thus producing the above rotation law. 
The star models are
essentially one--dimensional, which enormously simplifies
the computations.
Shellular rotation is likely to occur in  fast rotators
as well as in slow ones.  

The equations of stellar structure can be 
written consistently for a differentially rotating star, if  the
rotation law is shellular. Then, the isobaric surfaces  
satisfy the same equation $\Psi = const.$ as the equipotentials of the
conservative case (Meynet \& Maeder 1997) and thus the equations 
of stellar structure can be written as a function of the 
coordinate of an isobar, either in the lagrangian or eulerian form.
Let us emphasize that
in general the hydrostatic effects of rotation have only very small
effects of the order of a few percents on the internal evolution 
(Faulkner et al 1968; Kippenhahn \& Thomas 1970). Recent two--dimensional
models including the hydrostatic effects of rotation confirm the
smallness of these effects (Shindo et al 1997).

The above potential $\Psi$  
describes the shape of the star in the conservative case 
for the so--called Roche model, where the 
distorsions of the gravitational
 potential
are neglected. However, the stellar surface deviates from a surface
given by $\Psi=const.$ in case of non--conservative rotation law (Kippenhahn, 1977).

\subsection{THE VON ZEIPEL THEOREM}

The von Zeipel theorem (1924) is essential to predict
the distribution of temperature at the surface of a rotating star.
It applies to the conservative case and states that 
the local radiative flux
 $\vec{F}$ is proportional to the local effective gravity
$\vec{g_{\rm eff}}$, which  is the sum of the 
gravity and centrifugal force, if the star is not close to the
Eddington limit,

\begin{equation}
\vec{F} = - \frac{L(P)}{4\pi GM_\star(P)} \, \vec{g_{\rm eff}}
\end{equation}%eqn 1

\noindent with $M_{\star}(P) = M(1 - \frac{\Omega^2}{2\pi G\bar{\rho}})$;
$L(P)$ is  the luminosity on an isobar and $\bar{\rho}$ the mean
internal density.
Thus, the local effective temperature on the surface of a rotating
star  varies like
$T_{\rm eff}(\vartheta) \sim g_{\rm eff}(\vartheta)^{\frac{1}{4}}$. 
This  shows  that the spectrum of a rotating
star is in fact 
a composite spectrum made of local atmospheres of different gravity
and $T_{\rm eff} $. If it is meaningful to 
define an average, a reasonable choice
is to take $T_{\rm eff}^4= L/(\sigma S(\Omega))$, 
where $\sigma$ is Stefan's constant and $S(\Omega)$ the total actual
stellar surface. In the case of non--conservative rotation law, the corrections
to the von Zeipel theorem depend on the opacity law and 
on the degree of differential rotation, but they are likely to be small,
i.e. $\leq 1 \%$ in current cases of shellular rotation
(Kippenhahn 1977; Maeder 1999a).
There are some discussions (Heger et al 2000; Langer et al 1999)
whether von Zeipel must really be used or not. We would like to emphasize
that this is just a mere and unescapable consequence of
Newton's law and basic thermodynamics (Sect. 5.3).

\subsection{TRANSPORT OF ANGULAR MOMENTUM AND CHEMICAL ELEMENTS}

Inside a rotating star the angular momentum is transported by
convection, turbulent diffusion and  meridional circulation.
The equation of transport has been derived  by Jeans (1928),
Tassoul (1978), Chaboyer \& Zahn (1992), Zahn (1992). 
For shellular rotation, the equation of transport of angular
momentum in the vertical direction is (in lagrangian coordinates)

\begin{eqnarray}
\rho\frac{d}{dt}
\left( r^2 \Omega\right)_{M_{r}} = 
 \frac{1}{5 r^2}  \frac{\partial}{\partial r}
\left(\rho r^4 \Omega U(r)\right)
  + \frac{1}{r^2} \frac{\partial}{\partial r}
\left(\rho D r^4 \frac{\partial\Omega}{\partial r} \right) \;.
\end{eqnarray} %eqn.\ 2

\noindent $\Omega(r)$ is the mean angular velocity at level r,
$U(r)$  the vertical component of the meridional circulation
velocity and $D$  the  diffusion coefficient due to the sum
of the various turbulent diffusion processes (Sect. 2.5).
The factor $\frac{1}{5}$ comes
from the integration in latitude. If both $U(r)$ and $D$ are zero, we
just have the local conservation of the 
 angular momentum $r^2\Omega = const.$,
for a fluid element in case of contraction or expansion.
The solution of eqn.\ 2 gives the
\emph{non--stationary solution} of the problem.
 The rotation law is not arbitrary chosen, but
is allowed to evolve with time as a result of transport by meridional
circulation, diffusion processes and contraction or expansion.
 In turn, the 
differential rotation  built--up by these
 processes generates some turbulence and meridional circulation,
which are themselves functions of the rotation law.
 This coupling provides a feedback, and the  self--consistent
solution for the evolution of $\Omega (r)$ has to be found
 (Zahn 1992).

Some characteristic times can be associated,
to both the processes of meridional circulation and diffusion,

\begin{equation}
t_{\rm circ} \simeq \frac{R}{U} \hspace*{.2in},\hspace*{.5in}
t_{\rm diff} \simeq \frac{R^2}{D} \; . 
\end{equation} %eqn.\ 3

\noindent These timescales are essential quantities,
because the comparison with the nuclear timescales will show
the relative 
importance of the transport processes in the considered nuclear phases.
Equation 2 also  admits a \emph{stationary solution}, when one of 
the above  characteristic times  is short with respect to the
nuclear evolution time, a situation which only 
occurs at the beginning of the MS:

\begin{equation}
U(r)= - \frac{5 D}{\Omega} \frac{\partial \Omega}{\partial r} \; .
\end{equation}%eqn 4

This equation (Randers 1941; Zahn 1992) expresses that the (inward)
flux of angular momentum transported  by meridional circulation 
is equal to the (outward)
diffusive flux of angular momentum. As a matter of fact, 
this solution is equivalent to  considering local conservation of 
angular momentum (Urpin et al 1996).

Instead of  eqn.\ 2, the transport of
angular momentum by circulation 
is often treated as a diffusion process
(Endal \& Sofia 1978; Pinsonneault et al 1989, 1990; Langer 1991a;
Fliegner \& Langer 1995; Chaboyer et al 1995ab; Heger et al
2000). We see from eqn.\  2 that the term with $U$ (advection)
is functionally not the same as the term with  D (diffusion).
Physically advection and diffusion are quite different:
diffusion brings a quantity  from where there is a lot to other
places where there is little. This is not necessarily the case
for advection. As an example, the circulation of money in
the world is not a diffusive process, but rather an advective one !
Let us make clear that circulation with
a positive value of $U(r)$, i.e.\ rising 
along the polar axis and descending at the equator, 
is as a matter of fact making an \emph{inward} transport
of angular momentum. If this process were treated as a
diffusive function of $\frac{\partial \Omega}{\partial r}$, even
the sign of the effect may be wrong.

A differential equation like eqn.\ 2 is subject to boundary conditions
at the edge of the core and at the stellar surface. 
At both places, this condition 
is usually $\frac{\partial \Omega}{\partial r} = 0$,
with in addition the assumptions of solid body rotation for
the convective core 
and $U=0$ at the surface (Talon et al 1997; Denissenkov et al 1999).
If there is magnetic coupling at the surface 
(Hartmann \& Noyes 1987; Pinsonneault et al 1989, 1990; Pinsonneault 1997)
or mass loss by stellar winds,
the surface condition must be modified accordingly (Maeder 1999a).
 Various asymptotic regimes for the  angular momentum transport
can be considered  (Zahn 1992) depending on the presence of a wind 
with or without magnetic coupling.

\textsc{TRANSPORT OF CHEMICAL ELEMENTS}

The transport of chemical elements is also governed by a 
diffusion--advection equation like eqn.\ 2 (Endal \& Sofia 1978;
Schatzman et al 1981; Langer 1991a, 1992; Heger et al 2000).
However, if the
horizontal  component of the turbulent diffusion is large,
the vertical advection of the elements (and not that of the
angular momentum) can be treated as  a simple diffusion
(Chaboyer \& Zahn 1992) with a diffusion coefficient
$D_{\rm eff}$,

\begin{equation}
D_{\rm eff} = \frac{\mid rU(r) \mid^2}{30 D_h} \; ,
\end{equation} %eqn.\5

\noindent where $D_h$ is the coefficient of horizontal
turbulence, for which
the estimate is $D_h = |r U (r)|$ (Zahn 1992). Eqn.\ 5 
expresses that the vertical
advection of chemical elements is severely inhibited by the
strong horizontal turbulence characterized by $D_h$. 
 Thus, the change of the mass fraction $X_i$ of the chemical species 
\textit{i} is simply

\begin{equation}
\left(\frac{dX_i}{dt} \right)_{M_r} =
\left(\frac{\partial  }{\partial M_r} \right)_t
\left[ (4\pi r^2 \rho)^2 D_{\rm mix} \left( \frac{\partial X_i}
{\partial M_r}\right)_t
\right] + \left(\frac{d X_i}{dt} \right)_{\rm nucl} .
\end{equation} %eqn.\ 6

\noindent The second term on the right accounts for
 composition changes due
to nuclear reactions. The coefficient $D_{\rm mix}$ is the sum 
$D_{\rm mix} = D +D_{\rm eff}$,
where D is the term  appearing in expression 2 and 
$D_{\rm eff}$ accounts
for the combined effect of advection and horizontal 
turbulence. The characteristic time for the mixing
of chemical elements is therefore

\begin{equation}
t_{\rm mix} \simeq \frac{R^2}{D_{\rm mix}} \; .
\end{equation} %eqn.\ 7

Noticeably, the characteristic time for 
chemical mixing is not $t_{\rm circ}$   given by eqn.\ 3,
as has been generally considered (Schwarzschild 1958).
This makes the mixing of the chemical elements much slower,
since $D_{\rm eff}$ is very much reduced. 
In this context, we recall that several
authors have  reduced  by arbitrary factors, up to 30 or 100,
the effect of the transport
of chemicals in order to better fit the observed surface compositions
(Pinsonneault et al 1989, 1991; Chaboyer et al
1995ab; Heger et al 2000). This is no longer
necessary with the more appropriate expressions given above.

\subsection{MERIDIONAL CIRCULATION}

Meridional circulation arises from the local breakdown of radiative
equilibrium in a rotating star (Vogt 1925; Eddington 1925). In a
uniformly rotating
star, the equipotentials are closer to each other along the polar axis
than along the equatorial axis. Thus, according to von Zeipel's theorem,
the heating on an equipotential is generally 
higher in the polar direction than
in the equatorial direction, which thus drives a large scale
circulation rising at the pole and descending at the equator.
This problem has been studied for about 75 years (see reviews 
by Tassoul 1978 or Zahn 1983).
The classical formulation (Sweet 1950;
Mestel 1953, 1965; Kippenhahn \& Weigert 1990)
for rigid rotation  predicts a value of the vertical velocity
of the Eddington--Sweet circulation 

\begin{equation}
U_{ES} = \frac{8}{3} \omega^2
\frac{L}{gM}
\frac{\gamma  -1}{\gamma} \frac{1}{\nabla _{\rm ad} -\nabla}
\left( 1 - \frac{\Omega^2}{2\pi G\rho}\right) \; ,
\end{equation} %eqn.\8

\noindent with $\omega^2 = \frac{\Omega^2 r^3}{G M_r}$ the local ratio of 
centrifugal force to gravity and $\gamma$ the ratio of the specific
heats $C_P/C_V$.
The term $\frac{\Omega^2}{2\pi G\rho}$, often called the 
Gratton--\"{O}pik term (Tassoul 1990),
predicts that $U_{ES}$ becomes negative at the stellar surface
due to the presence of the term $1/\rho$. This 
means an inverse circulation, i.e. descending
at the pole and rising at the equator. The dependence in 
$\frac{1}{\rho}$ also makes $U(r)$ diverge at the surface. This has 
led to some controversies on what is limiting $U(r)$
(Tassoul \& Tassoul 1982, 1995;
Zahn 1983; Tassoul 1990).
The timescale for circulation mixing defined in eqn.\ 3 becomes
with the above Eddington--Sweet velocity 

\begin{equation}
t_{\rm ES} \simeq t_{\rm KH} \frac{g}{\Omega^2 R} \; ,
\end{equation} %eqn.\ 9

\noindent where $g$ is the surface gravity and $R$ the 
stellar radius. Even for modest
rotation velocities, $t_{\rm ES}$ is much shorter than the MS lifetime
(Schwarzschild 1958; see also Denissenkov et al 1999),
so that most stars should be mixed, if this timescale were applicable.
However, the presence of $\mu$--gradients 
was not taken into account in the above expressions. 
When this is done, rotation is found
to allow circulation only above a certain
rotation limit that depends on the value of
the $\mu$--gradient (Mestel 1965;
Kippenhahn 1974; Kippenhahn \& Weigert 1990). 

The velocity of the meridional circulation in the case of shellular 
rotation was derived by Zahn (1992), who considered 
the effects of the  latitude--dependent $\mu$--distribution
(Mestel 1953, 1965).
Even more important are the effects of the vertical
$\mu$--gradient  $\nabla_{\mu}$  and of the horizontal 
turbulence (Maeder \& Zahn 1998).  
Contrary to the conclusions of the previous works,  
the $\mu$--gradients  were shown not to introduce a
velocity threshold for the existence of the circulation,
 but to progressively reduce
 the circulation when $\nabla_{\mu}$ increases. One has then    

\begin{eqnarray}
U(r) &=&  \frac{P}{{\rho} {g} C_{\!P} {T}
\, [\nabla_{\rm ad} - \nabla +  (\varphi/\delta) \nabla_{\mu}] } 
  \left\{  \frac{L}{M_\star} (E_\Omega + E_\mu) \right\} \; ,
\end{eqnarray} %eqn.\ 10

\noindent where P is the pressure, $C_P$ the specific heat, 
$E_{\Omega}$ and $E_{\mu}$ are  terms depending on the $\Omega$--
and $\mu$--distributions respectively, up to the third order derivatives.
Since the derivative of $U (r)$ appears in eqn.\ 2, we see that
the consistent solution of the problem  is of fourth order
(Zahn 1992). This makes the numerical solution
difficult  (Talon et al
1997; Denissenkov et al 1999; Meynet \& Maeder 2000). 
While the classical solution predicts an infinite
velocity at the interface between a radiative and a semiconvective
zone with an inverse circulation in the semiconvective
zone, expression 10
gives a continuity of the solution with no change of sign.
In evolutionary models, the term  $\nabla_{\mu}$ in eqn.\ 10 may be
one or two orders of magnitude larger than $\nabla_{\rm ad}-\nabla$
in some layers,
so that $U (r)$ may be reduced by the same ratio. This considerably
increases the characteristic time $t_{\rm circ}$
with respect to the classical estimate $t_{\rm ES}$.

\subsection{INSTABILITIES AND TRANSPORT}

The subject of the instabilities in moving plasmas is a
field in itself. Here we limit ourselves to a short
description of the main instabilities 
currently considered influential in the evolution 
of Upper MS stars, (Endal \& Sofia 1978; Zahn 1983, 1992;
Pinsonneault et al 1989; Heger et al 2000).

\textsc{CONVECTIVE AND SOLBERG-H\mbox{\O}ILAND INSTABILITY}

In a rotating star, the Ledoux or Schwarzschild criteria for
convective instability should be replaced  by the 
Solberg--H\mbox{\o}iland
criterion (Kippenhahn \& Weigert 1990). This criterion accounts 
for the difference of the centrifugal
force for an adiabatically displaced  fluid element; the condition
for dynamical stability is

\begin{equation}
N^2 +
\frac{1}{s^3} \frac{d(s^2 \Omega)^2}{ds} \geq 0 \; .
\end{equation} %eqn.\ 11

\noindent The Brunt-Väisälä frequency $N^2$ is given by 
$N^2 = \frac{g \delta}{H_p} \left( \nabla_{\rm ad}-\nabla +
 \frac{\varphi}{\delta} \nabla_{\mu} \right)$, where the various symbols
have their usual meaning (Kippenhahn \& Weigert 1990), and the term 
$s$ is the distance to the rotation axis. For no rotation,
the Ledoux criterion is recovered. If the thermal effects are
ignored, we just recover Rayleigh's criterion for stability, which says
that the specific angular momentum $s^2\Omega$ must increase with
the distance to  the rotation axis.
In practice,  convective stability is reinforced by rotation.
In the absence of rotation, a zone located beween the places where
$\nabla = \nabla_{\rm ad} + \nabla_{\mu}$ and 
$\nabla = \nabla_{\rm ad}$ is called  semiconvective.
There, non--adiabatic effects can drive
 growing oscillatory instabilities (Kato 1966; Kippenhahn \& Weigert
1990).
An appropriate  diffusion coefficient  describing the transport
in such zones has been derived by Langer et al (1983). However
there is no diffusion coefficient  yet available for semiconvective
mixing in the  presence of rotation.

The assumption of solid body rotation is generally
made in convective regions,
owing to the strong turbulent coupling. However, 
the collisions or scattering  of convective blobs influence the 
rotation law in convective regions (Kumar et al 1995): solid body
rotation only occurs for an isotropic scattering. For some
forms of anisotropic
scattering, an outward
rising rotation profile such as that observed in the Sun, can be produced.
Two--dimensional models of rotating stars (Deupree 1995, 1998)
also show that the angular velocity  in convective cores 
is not uniform, but it decreases with distance from the
center, being about constant on cylinders. A considerable 
overshoot is  obtained by Deupree (1998), amounting to about  0.35 $H_P$,
where $H_P$ is the local pressure scale height.
Similar conclusions are obtained by Toomre (1994), who also finds
penetrative convection at the edge of the
convective core of rotating A--type stars. 
The very large Reynolds number characterizing stellar turbulence
prevents direct numerical simulations of the Navier--Stokes equation,
so that  some new specific methods have been proposed to study 
convective turbulence (Canuto 1994; Canuto et al 1996;
Canuto \& Dubovikov 1997) and its interplay
with differential rotation (Canuto et al 1994; Canuto 1998); these
last results have not yet been applied to evolutionary models in rotation.

\textsc{SHEAR INSTABILITIES: DYNAMICAL AND SECULAR}

In a radiative zone, shear due to differential rotation is likely
to be a very efficient mixing process. Indeed shear instability grows on a dynamical
timescale that is of the order of the rotation period 
(Zahn 1992, 1994). Stability is maintained when the Richardson 
number $Ri$ is larger than a critical value $Ri_{\rm cr}$
 
\begin{equation}
Ri = \frac{N^2}{\left(\frac{dV}{dz}\right)^2} > 
Ri_{\rm cr} =\frac{1}{4} \; ,
\end{equation} %eqn.\\ 12

\noindent where $V$ is  the horizontal
 velocity and $z$ the vertical coordinate. 
Equation 12 means that the restoring force of the density
gradient
is larger than the excess energy $\frac{1}{4}\left(\frac{dV}{dz}
\right)^2$ present  in the differentially rotating layers
(Chandrasekhar 1961). In eqn.\ 12, heat exchanges are ignored
and the criterion refers to the ``dynamical shear instability''
(Endal \& Sofia 1978; Kippenhahn \& Weigert 1990).

 When thermal dissipation is 
significant, the restoring force of buoyancy is reduced and
the instability occurs more easily (Endal \& Sofia 1978), 
its timescale is however longer, being the thermal timescale.
This case is sometimes referred to as ``secular shear
instability''. For small thermal effects ($Pe \ll 1$), a 
factor equal to $Pe$ appears as multiplying 
$N^2$ in eqn.\ 12 (Zahn 1974). The number  $Pe$
is the ratio of the thermal cooling time to the 
dynamical time, i.e.  $Pe = \frac{v \ell}{K}$ 
where $v$ and $\ell$ are the 
characteristic velocity and  length scales, and $K = (4acT^3)/
(3 C_P \kappa \rho^2 )$ is the thermal diffusivity. 
$Pe$  varies typically from $10^{9}$ in deep 
interior to $10^{-2}$ in outer layers (Cox \& Giuli 1968).
For general values of $Pe$, a more general expression of the 
Richardson's criterion  can be found (Maeder 1995;
Maeder \& Meynet 1996); it is consistent with the case of low $Pe$
treated by Zahn (1974). The problem of the Richardson criterion
has also been 
considered by Canuto (1998), who suggests that for $Pe > 1$, i.e.
negligible radiative losses $Ri_{\rm cr} \sim 1 $ and for 
$Pe < 1$, i.e. important radiative losses
$Ri_{\rm cr} \sim Pe^{-1}$. Thus, similar dependences with
respect to $Pe$ are obtained, but Canuto (1998)
finds that turbulence may exist beyond 
the $\frac{1}{4}$ limit in eqn.\ 12.

Many authors have shown that the 
$\mu$--gradients  appear to inhibit  the mixing
too much with respect to what is required by the observations
(Chaboyer et al 1995ab; Meynet \& Maeder 1997; Heger et al 2000).
Changing $Ri_{\rm cr}$ from $\frac{1}{4}$ to $1$ does not solve
the problem, the difference being a matter of 
one or two orders of  magnitude. For example, instead of
using a gradient $\nabla_{\mu}$, some authors write
 $f_{\mu} \nabla_{\mu}$ with a factor $f_{\mu} = 0.05$ 
or even  smaller (Heger et al 2000; see also Chaboyer et al 1995ab).
Most of the zone  external to
the convective core, where the $\mu$--gradient
inhibits mixing, is as a matter of fact semiconvective
and is thus subject to thermal instability anyway. 
This has led to the 
hypothesis that the excess energy in the shear
is degraded by turbulence on the thermal timescale, 
changing the entropy gradient 
and consequently the $\mu$--gradient (Maeder 1997). This
gives  a diffusion coefficient $D_{\rm shear}$, which
 tends towards the diffusion coefficient for semiconvection
by Langer et al (1983)
when shear is negligible, and towards the value
$D_{\rm shear} = (K/N^2) (\Omega \frac{d\ln\Omega}{d\ln r})^2$
given by Zahn (1992)
when semiconvection is negligible.  Another proposition
has been made by Talon \& Zahn (1997), who take into account
the homogeneizing effect of the horizontal diffusion on the
restoring force produced by the $\mu$--gradient. This also
reduces the  excessive stabilizing effect of the $\mu$--gradient.
Both the above suggestions lead
to an acceptable amount of mixing in view of the observations.
We stress that the Reynolds condition 
$D_{\rm shear} \geq \frac{1}{3} \nu Re_c$ 
must be satisfied when the medium is turbulent, where
 $Re_c$ is the critical Reynolds number (Zahn 1992; 
Denissenkov et al 1999) and $\nu$ is the  total viscosity
(radiative + molecular).

Globally,  we may expect the secular instability to work during
the MS phase, where the $\Omega$--gradients are small and the lifetimes
are long, while the dynamical instability could play a role
in the advanced stages.

\textsc{OTHER INSTABILITIES}

Baroclinicity, i.e.\ the non--coincidence of the equipotentials 
and surface of constant $\rho$, generates various  instabilities
in the case of non--conservative rotation laws. Some instabilities are  
axisymmetric, like the GSF instability (Goldreich \&
Schubert 1967; Fricke 1968; Korycansky 1991).
The GSF instability is created by
fluid elements displaced between the directions  of
constant angular momentum and of the rotational axis. 
Stability demands a uniform or constant rotation on cylinders, which is incompatible
with shellular rotation. The GSF instability
thus favours solid--body rotation, on a time--scale of the
order of that of the meridional circulation (Endal \& Sofia 1978). 
This instability is however  inhibited by the $\mu$--gradients
(Knobloch \& Spruit 1983), nevertheless Heger et al (2000) find it to 
play a role near the end of the helium--burning phase.
Another axisymmetric instability is 
the ABCD instability (Knobloch \& Spruit 1983).  Fluid
elements displaced between the surfaces of constant $P$ and $T$ 
create the ABCD instability, a kind of horizontal convection.
The ABCD instability is an oscillatory one and its efficiency 
is difficult to estimate for now.
  Non--axisymmetric instabilities,
like salt--fingers, may also occur (Spruit \& Knobloch 1984).
They are not efficient when rotation is low.
 However in the case of fast
rotation they may occur everywhere in rotating
stars, so that one--dimensional models are likely to be
an unsatisfactory idealization in this case.

The study of the transport of angular momentum 
by gravity waves has been stimulated 
by the finding of an almost solid body rotation for most of the
radiative interior of the Sun (Schatzman 1993; Montalban 1994;
Kumar \& Quataert 1997; Zahn et al 1997; Talon \& Zahn 1998).
Gravity waves  are supposed  to transport angular momentum
from the external convective layers to the radiative interior. 
However, Ringot (1998) has recently shown that quasi--solid rotation 
of the radiative zone of the Sun cannot be a direct consequence
of the action of gravity waves. Thus, even in the Sun the 
question remains open.

 In Upper MS stars, 
we could expect gravity waves to be generated by turbulent
motions in the convective core (Denissenkov et al 1999).
The momentum will be deposited  where the Doppler
shift of the waves due to differential rotation is equal 
to the initial wave frequency. From the work by
Montalban \& Schatzman (1996),  we know that in general 
the deposition of energy  decreases very quickly away from
the boundaries of a convective zone.  The same is found by
Denissenkov et al (1999), who show that uniform rotation
sustained by gravity waves is limited to the very inner 
radiative envelope; the size of the region of uniform
rotation enforced by gravity waves likely increases with 
stellar mass. Only angular momentum may be  directly transported 
by gravity waves and not
chemical elements.
Nevertheless, the transport of momentum  by waves,
which reduces the  differential rotation, could also influence 
indirectly the distribution of chemical elements in stars. 

\subsection{MASS LOSS AND ROTATION}

Mass loss by stellar winds is a dominant effect in the 
evolution of Upper MS stars (Chiosi \& Maeder 1986). The mass
loss rates currently applied in stellar models 
are based on the observations (de Jager et al 1988;
Lamers \& Cassinelli 1996). A  significant growth of the mass flux
of OB stars with rotation, i.e.\ by 2--3 powers of 10, was found
by Vardya (1985).  Nieuwenhuijzen \& de Jager (1988)
suggested that the correlation found by Vardya mainly reflects
the distributions of the mass loss rates $\dot{M}$ and of the 
rotation velocities $v_{\rm rot}$ over the HR diagram.
After trying to disentangle the effects of $L$, $T_{\rm eff}$ and
$v_{\rm rot}$, they found that the $\dot M$-rates 
seem to increase only slightly
with rotation for O-- and B--type stars. The result by
 Vardya might not be
incorrect, since when the data for OB stars
 by Nieuwenhuijzen \& de Jager (1988)
are considered, a correlation of the mass fluxes with $v_{\rm rot}$
is noticeable.
These authors also point out 
that the equatorial $\dot{M}$--rates of Be--stars
are larger by a factor $10^2$. Since Be--stars are essentially
B--stars with fast rotation,  a 
single  description of the large changes of the  
 $\dot{M}$--rates from the low to the high  values of $v_{\rm rot}$ 
should be considered.

 On the theoretical side, 
Pauldrach et al (1986) and Poe \& Friend (1986) find a very
weak change of the $\dot{M}$--rates with 
$v_{\rm rot}$ for O--stars: the increase amounts to about 30\%
for $v_{\rm rot}$ = 350 km/sec. 
Friend \& Abbott (1986) find an increase of the $\dot{M}$--rates
which can be fitted by the relation (Langer 1998;
 Heger \& Langer 1998)

\begin{equation}
\dot{M}(v_{\rm rot})\; = \;\dot{M} (v_{\rm rot} = 0) 
\left(\frac{1}{1-\frac{v_{\rm rot}}
{v_{\rm crit}}}\right)^{\xi}
\end{equation} %eqn.\ 13

\noindent with $\xi$ = 0.43; this 
expression is often used in evolutionary models.

The previous wind models of rotating stars are incomplete 
since they do not account for the von Zeipel theorem. The gravity
darkening at the equator leads to a reduction of
the equatorial mass flux (Owocki et al 1996; Owocki \& Gayley
1997, 1998).
This leads   to very different predictions for the wind morphology
than those of the current wind--compressed
disk model by Bjorkman \& Cassinelli (1993), 
which is currently advocated to explain disk formation.
 Equatorial disks may however form quite naturally around 
rotating stars.
The theory of radiative winds, with revised expressions
of the von Zeipel theorem and of the Eddington factor, has
been applied to rotating stars (Maeder 1999a). There are two main
sources  of wind anisotropies: 1. The ``$g_{\rm eff}$--effect''
which favours polar ejection,  since the polar caps of a rotating
star are hotter. 2. The ``opacity or $\kappa$--effect'', which favours 
an equatorial ejection, when the opacity is large enough at the equator
due to an opacity law which increases rapidly with
decreasing temperature. In O--type stars, since opacity is due 
mainly to the T--independent  electron scattering,  
the $g_{\rm eff}$--effect
is likely to  dominate and to  raise a fast highly ionized
polar wind.
 In B-- and later type stars,
where a T--growing opacity is present in the external layers, the
opacity effect should favour a dense equatorial wind and ring formation,
with low terminal velocities and low ionization.

The so--called B[e] stars (Zickgraf 1999) are known to show both 
a fast, highly ionized polar wind
and a slow, dense, low ionized equatorial ejection and may be
a template of the $g_{\rm eff}$--  and $\kappa$--effects.
At some values of T, the ionization equilibrium of the stellar wind
changes rather abruptly and so does the opacity as well as the 
force--multipliers which characterize the opacities
(Kudritzki et al 1989; Lamers 1997; Lamers 1999). Such transitions,
 called the bi--stability of stellar winds by Lamers,
may favour strong  anisotropies of the winds, and even create 
some symmetrical rings at the latitude where an opacity--peak
occurs on the T--varying surface of the star. 

It could be thought at first sight that wind anisotropies have
no  direct consequences for stellar evolution. This is not
at all the case.  Like magnetic coupling for low mass stars,
the anisotropic mass loss removes selectively the angular
momentum (Maeder 1999a) and influences the further evolution. 
Winds  through polar caps, as likely in O--stars,
remove very little angular momentum, while equatorial mass
loss removes a lot of angular momentum from the stellar 
surface. 

\section{MAIN SEQUENCE EVOLUTION OF ROTATING STARS}

\subsection{EVOLUTION OF THE INTERNAL ROTATION}

 As a result of transport  processes, contraction and expansion, 
the stars should be  differentially 
rotating,  with a
 strong horizontal turbulence
enforcing  a rotation law of the form $\Omega$ constant on isobars (Zahn 1992).
The whole problem must be treated self--consistently, because 
 the differential rotation in turn determines the behaviours
of the  meridional circulation
and turbulence, which themselves contribute to differential rotation.
There are various
approximations to treat the above physical problem
(Pinsonneault et al 1989, 1991; Chaboyer et al 1995ab;
Langer 1998; Heger \& Langer 1998; Heger et al 2000).
In some works, the assumption of rigid rotation is made, while
in  other works advection is treated 
as a diffusion  with the risk that even the sign of the effect
is the wrong one! Some authors, in order to fit the observations,
 introduce several efficiency factors
$f_{\mu},\; f_c$, etc... The problem is that 
the sensitivity of the results to these  many efficiency factors is 
as large, or even larger than  the sensitivity to rotation.

A simplification has been applied by Urpin et al (1996),
who assume equilibrium between the outward transport
of angular momentum by diffusion and the inward transport by circulation,
as also suggested by Zahn (1992).
This is the stationary case discussed in Sect.\ 2.3 for which eqn.\ 4
applies. The values of $U(r)$ are always positive 
and the circulation has only one loop.
Urpin et al (1996) point out  that the stationary distribution
of $\Omega$ arranges itself so as to reduce $U(r)$ to a minimum
value over the bulk of the star. 
This makes the values 
of $U(r)$ of the order of $10^{-5} - 10^{-6}$ cm/s, 
quite insufficient to produce any efficient mixing
in a 20 M$_{\odot}$ star. 

The initial non--stationary approach to equilibrium
 in 10 and 30 M$_{\odot}$ stars
has been studied by Denissenkov et al (1999). In a very
short time of about 1\% of the MS lifetime $t_{MS}$,
 $\Omega(r)$ converges towards a profile with 
a small degree of differential rotation and
with very small values of $U(r)$ (Urpin et al 1996).
 The circulation shows two cells,
an internal one rising along the polar axis and an external 
one descending  at the pole. The evolution is not calculated,
but it is noted that
the timescale $t_{\rm circ}$ (which behaves like 
$\Omega^{-2}$) is very  short  with respect to 
the MS lifetime for most Upper MS stars and that 
the ratio $\frac{t_{\rm mix}}{t_{MS}} \geq 1$ (eqn.\ 7).
It is noticeable that $t_{\rm circ}  \ll t_{\rm mix}$,
so that no efficiency factors are needed to reduce the mixing of 
chemical elements compared to the transport of angular momentum.
This reduction naturally results from the 
effect of the horizontal turbulence.

The  full evolution of the rotation law has been studied 
with the non--stationary  scheme for a 9 M$_{\odot}$ star
by Talon et al (1997), and for stars from 5 to 120 M$_{\odot}$
including the effects of mass loss by
Meynet \& Maeder (2000). The very fast initial convergences of 
$\Omega(r)$ and of $U(r)$ are confirmed. However, after
convergence  the asymptotic
state of $U(r)$ does  not correspond 
to the stationary approximation.
In the full solution, $U(r)$ changes sign in the external region
and thus transports some angular momentum outward,
which  is not the case in the stationary solution.
Also, contrary to the classical result of the 
Eddington--Sweet circulation  (eqn.\ 8), it is
found  that $U(r)$ depends very little on the initial rotation.

Fig.\ 1 shows the evolution of $\Omega(r)$  during
MS evolution of a 20 M$_{\odot}$ star.
Mass loss at the stellar surface removes a substantial
fraction of the total angular momentum, which makes 
 $\Omega(r)$ decrease with time everywhere in the star.
 The outer zone 
with inverse  circulation progressively deepens during MS evolution,
because a growing part of the outer layers has  lower densities.
This inverse circulation contributes to the outward transport
of angular momentum. The deepening
of the inverse circulation also has the consequence that the
stationary and non--stationary solutions differ more and
more as the evolution proceeds, since  no inverse
circulation is predicted by the stationary solution.
This shows that the stationary
solutions  are too simplified and that outward and inward 
transport never reach exact equilibrium, contrary to the initial
expectations. 

Fig.\ 2 shows the various diffusion coefficients inside a 20 
M$_{\odot}$ star when the hydrogen mass fraction at the center is 
equal to 0.20.
We notice that in general $ K \geq D_h \geq D_{\rm shear} \geq
D_{\rm eff}$. This confirms  the basic hypothesis 
of a large  $D_h$ necessary for the validity of
the assumption of shellular rotation.  

The characteristic time 
$t_{\rm mix}$ of the mixing
processes is of the same order  as the lifetime $t_{\rm H}$ of the
H--burning phase for the Upper MS stars (Maeder 1987). 
Indeed, if shear mixing is the dominant mixing  process,
the timescale is $t_{\rm mix} \simeq 
\frac{R^2}{D_{\rm shear}}$, and for a given degree of differential
rotation it behaves like $t_{\rm mix} \simeq  K^{-1}$, 
which itself goes like about $M^{-1.7}$. The timescale  $t_{\rm H}$
behaves like $M^{-0.7}$ for $M \geq 15$ M$_{\odot}$ (Maeder 
1998). Thus, for larger masses $t_{\rm mix}$  tends to decrease 
 much  faster than $t_{\rm H}$ and mixing processes 
grow in importance.
From  the end of MS evolution when $X_c \leq 0.05$, central contraction
starts dominating the evolution of the central $\Omega(r)$, which
grows quickly until core collapse.
During these post--MS phases, the average value of
$t_{\rm mix}$ will be longer 
than the nuclear lifetimes. Thus, the rotational  mixing 
processes during these phases are likely  to be globally unimportant 
(Heger et al 2000; Meynet \& Maeder 2000).
 Nevertheless, it
 is likely that in some
regions of a  rotating star in the advanced stages, 
as a result of extreme central contraction, the $\Omega$--gradient
may become so large that some other local instabilities develop 
leading to fast mixing.

\subsection{EVOLUTION OF $V_{\rm rot}$. THE CASE OF Be STARS.}

The evolution of the surface rotational velocities 
$v_{\rm rot}$ at the equator is
a consequence of the processes discussed above. Let us 
first consider the two extreme cases of coupling
and of no coupling between adjacent layers, first examined
in the early works by Oke \& Greenstein (1954) and
Sandage (1955): a) the case of rigid rotation and b)
the case of local conservation. a) For rigid rotation, 
$v_{\rm rot}$ remains nearly constant during the MS phase
(Fliegner \& Langer 1995). This is because
the effects of core contraction and envelope expansion
nearly compensate for one another. b) For
local conservation of the angular momentum, one has 
$v_{\rm rot}= \Omega \:R \sim  R^{-1}$, while the critical velocity
changes like  $v_{\rm crit} \sim R^{-\frac{1}{2}}$. 
Thus, the inflation  of the stellar radius $R$ during evolution 
makes rotation less and less
critical. The opposite effect may occur during a
bluewards crossing of the HR diagram and then critical rotation
may be reached  (Sect. 5.2). Endal \& Sofia (1979)
have shown that the ratio 
 $v_{\rm rot}(case \;a)/v_{\rm rot}(case \;b) \simeq 1.8$
before the crossing of the HR diagram for intermediate mass stars.
 The truth generally lies
 between the two cases, closer to the
rigid case during the MS phase since transport processes have more 
time to proceed. In post--MS phases, and in particular during
the fast crossings  of the HR diagram, the evolution 
timescale is short, 
 so that  little transport occurs and the evolution of 
$v_{\rm rot}$  closely resembles that of local conservation.

Stars in solid body rotation  may reach the break--up
limit before the end of the MS phase
even for moderate initial $v_{\rm rot}$ (Sackmann \& Anand 1970;
Langer 1997). However this is a consequence of the simplified
assumption of solid body rotation. With diffusion and transport,
it is less easy  for the star to reach the break--up limit.
For a 20 M$_{\odot}$ model without mass loss (dotted line in 
Fig. 3 and 4), $v_{\rm rot}$  grows 
and the ratio $\frac{\Omega}{\Omega_{\rm crit}}$ may
become close to 1 before the end of the MS phase. 
For stars with M $<$ 15 M$_{\odot}$, where mass loss is small,
the ratio $\frac{\Omega}{\Omega_{\rm crit}}$ also increases during
the  MS phase  and the break--up or $\Omega$--limit may be 
reached during the MS phase.  If so,
the mass loss should then become very intense, 
until the velocity again becomes subcritical. It is somehow 
paradoxical that no or small mass loss rates
during the bulk of the MS phase (as at low metallicity Z)
may lead to very high mass loss
at the end of the MS evolution for rotating stars.
This may be the cause of some ejection processes as
in Luminous Blue Variables (LBV), B[e] and Be stars. We may 
wonder whether the higher relative number of Be stars observed
in lower Z regions (Maeder et al 1999) is just
a consequence of  the  lower average mass loss
in lower metallicity regions or whether this is 
related to star formation.

If mass loss is important during MS evolution, $v_{\rm rot}$
decreases substantially (Fig. 3 and 4).
 Therefore it is not surprising that Be--stars, which
likely are close to break--up (Slettebak 1966)
do not form  among O--type stars, but mainly among B--type stars,
with a relative  maximum at type B3. 
Indeed, to form a Be star it is probably not necessary that the 
break--up limit is exactly reached.
The conditions for an equatorial ejection responsible
for  the Be spectral features  occur when  the $\kappa$--effect is
important (Sect. 2.6), which requires that the equatorial
 regions of the fastly 
rotating star are below the bi--stability limit.
Of course, the higher the rotation, the
higher the equatorial mass loss will be. 

Any magnetic coupling between the star and the wind
would dramatically reduce $v_{\rm rot}$. 
However, such a coupling 
 does not seem to be important in general, except for Bp and Ap stars.
 MacGregor et al (1992)
show that, even in the presence of a small magnetic field 
of 100 G,
the rotation velocities of OB stars
should be much lower than observed. This result is in
agreement with that of Mathys (1999), who finds no detectable 
magnetic field  in hot stars.

\textsc{EFFECTS OF MASS LOSS ON ROTATION} 

Mass loss by stellar winds  drastically reduces 
 $v_{\rm rot}$ during the evolution (Packet et al 1980; 
Langer \& Heger 1998; Figs. 3 and 4). Even if isotropic,
the stellar winds carry away quite
 a lot of angular momentum, and this 
is even  more important in the case of equatorial mass loss.
The new surface layers then have a lower $v_{\rm rot}$ as 
a result of expansion and redistribution. 

With the simplified assumption of solid body rotation
 for a 60  M$_{\odot}$ model with mass loss, Langer (1997, 1998)
finds a convergence of $v_{\rm rot}$ 
towards the critical value (the $\Omega$--limit),
 before the end of the MS phase,
 for all initial velocities  above
100 km/s. The overall result is that the final velocities
are the same, all being  at the critical limit, while
 the final MS masses strongly depend on the
initial rotation.
This convergence towards 
the  $\Omega$--limit clearly 
results also from the  simplified assumption of rigid rotation,
which grossly exagerates the coupling of the surface layers.

Fig. 3 illustrates
the decrease in $v_{\rm rot}$ during evolution when
the various transport mechanisms are followed in detail.
The  decrease  in $v_{\rm rot}$ is much larger
for larger initial stellar masses, because mass loss is larger
for them. The same is true for $\frac{\Omega}{\Omega_{\rm crit}}$
(Fig. 4). For $M \geq 40$ M$_{\odot}$
the  velocities $v_{\rm rot}$ will remain largely subcritical
for all initial velocities, except during the overall
contraction phase at the end of MS evolution.
For a given initial mass, 
the resulting scatter of $v_{\rm rot}$ should be 
smaller at the end 
of the MS phase, but this is not a convergence 
towards the $\Omega$--limit  as for models 
with solid body rotation (Langer 1997, 1998).
 The final masses are of course lower for
larger rotation, since  mass loss is enhanced by rotation, this
 results in a large scatter of masses and $v_{\rm rot}$ 
at a given luminosity.

\textsc{COMPARISON WITH OBSERVATIONS}

Only a few  comparisons between the observed $v_{\rm rot}$ and the model
predictions  have been made until now. Conti \& Ebbets (1977) 
found that $v_{\rm rot}$ in O--type giants and supergiants
is not as low as expected from models with rotation
conserved in shells. This fact and the relative absence of low 
rotators among the evolved O--stars led them to conclude that another
line broadening mechanism, such as macroturbulence, should be present
in these objects. The same conclusion based on similar
arguments was supported by Penny (1996) and Howarth et al (1997).
This conclusion needs to be  checked further, since  the 
decrease predicted by the new models  is not as fast as 
for local conservation.

It is well--known that the average value of  $v_{\rm rot}$ 
increases from the early O--type to the early B--type stars 
(Slettebak 1970). 
This may be the signature of the effect of
higher losses of mass and angular momentum in the more
massive stars (Penny 1996), which leads to a lower  average rotation
 in the course of the MS phase. We notice that the increase 
of $v_{\rm rot}$ from O to B stars is 
larger for the  stars of class IV  than for the stars of class V
(Fukuda 1982),
a fact which is expected since the difference due to mass loss
is more visible near the end of the MS phase (Fig. 3).
The stars of luminosity class I (Fukuda 1982) show a fast decline 
of the average $v_{\rm rot}$ from the O--type to the B--type stars.
As the supergiants of class I originate from the most massive
stars, which evolve at about constant luminosity,
this  last effect could be  related 
to the fact that for a given initial mass, 
$v_{\rm rot}$  declines strongly as the star moves
away from the MS (Langer 1998). 

\subsection{THE HR DIAGRAM, LIFETIMES AND ISOCHRONES}

As always in stellar evolution, the shape of the tracks is closely
related to the internal distribution of the mean molecular weight
$\mu$. All results show  that 
the convective cores are slightly increased by rotation
(Maeder 1987, 1998; Langer 1992; Talon et al 1997;
Meynet 1998, 1999; Heger et al 2000). The height of the 
$\mu$--discontinuity at the edge of the core is reduced
and  there is a mild composition gradient built 
up from the core to the surface, which may then be slightly
enriched in helium  and nitrogen.
  For low or moderate rotation, the
convective core shrinks  as usual during MS evolution,
while for high masses ($M \geq 40$ M$_{\odot}$) and  large initial rotations 
($\frac{\Omega}{\Omega_{\rm crit}}
\geq 0.5 $), the convective core grows in mass  during 
evolution. These behaviours, i.e.\ reduction or growth of the core,
determine  whether the star will follow 
respectively the usual redwards MS tracks in the HR diagram, 
or whether it will bifurcate to the blue (cf. Maeder 1987; Langer 1992)
towards the classical tracks of homogenous evolution (Schwarzschild 1958)
and likely produce W--R stars (Sect. 5.4).
Also for O-- and B--type stars, fast rotation increases 
the He--content of the envelope and the decrease of the
opacity also favours a bluewards track. 

Fig. 5 shows the overall HR diagram for rotating and non--rotating
stars during the MS phase and slightly beyond. The atmospheric
distortions produce a shift 
to the red in the HR diagram by several 0.01 in (B-V),
with on the average only a small change of luminosity
(Maeder \& Peytremann 1970; Collins \& Sonneborn 1977).
During MS evolution, the luminosity of the rotating stars  grows
faster and the tracks extend farther away from the ZAMS, 
as in the case of a moderate overshooting (Maeder 1987;
Langer 1992; Sofia et al 1994; Talon et al 1997).
This effect introduces a significant scatter in the mass--luminosity
relation (Meynet 1998), in the sense that fast rotators are
overluminous with respect to their actual masses.
This may explain some of the discrepancies
between the evolutionary masses and the direct
mass estimates in some
binaries (Penny et al 1999). In this context, 
we recall that for a decade a severe
mass discrepancy beween spectroscopic and evolutionary masses
 was claimed by some authors (Groenewegen et al
1989; Kudritzki et al 1992; Herrero et al 1992). Most of the
problem has collapsed and was shown 
 to be a result of the proximity of O--stars to the Eddington limit 
(Lamers \& Leitherer 1993; Herrero et al 1999) and the large effect
of metal line blanketing not usually accounted for in the
atmosphere models of massive stars (Lanz et al 1996).

There is little difference between tracks with 
$v_{\rm rot}$ = 200 or 300 km/s (Meynet \& Maeder 2000; see also Talon et al 1997).
If the effects behaved  like $v_{\rm rot}^2$, there 
would be  larger differences.
This saturation effect occurs because outward transport of angular momentum by shears are larger when rotation is larger, also
a larger rotation produces 
more mass loss, which further reduces rotation during the evolution. 

\textsc{LIFETIMES AND ISOCHRONES}

The lifetimes $t_{\rm H}$ in the H--burning phase
grow  only moderately 
because more nuclear fuel is available, but at the same
time the luminosity is larger. The net result is an increase 
 by about 20 to 30 \% for an initial velocity of 200 km/s
(Talon et al 1997; Meynet \& Maeder 2000). This influences
the isochrones and the  age determinations. As an example, 
for $v_{\rm rot}$ = 200 km/s, the isochrone of log age = 7.0 
is  the same  as that of log age = 6.90 without rotation
(Meynet 2000). Thus, accounting for rotation could lead to
ages larger by about 25 \% for O-- and early B--type
stars. However, since the cluster ages are generally
determined on the basis of the blue envelope of the
observed  sequence,  where most low rotators lie 
(Maeder 1971), it is likely that the effect in current
age determinations is rather small.
If a bluewards track occurs, the larger core and mixing lead to
much longer lifetimes in the H--burning phase.
In this case, the fitting of time--lines becomes hazardous. 

\section{ROTATION AND CHEMICAL ABUNDANCES}

\subsection{OBSERVATIONS}

The chemical abundances are a very powerful test of internal
evolution and they give strong evidence in favour of
some additional mixing processes in O-- and B--type stars,
in supergiants and in red giants of lower masses.
 
\textsc{ ABUNDANCES IN MS  O--AND B--TYPE STARS}

Many evidences of  He-- and N--excesses in
O--type and early B--type stars have been reported over the last decade
(Gies \& Lambert 1992; Herrero et al 1992, 1998; Kilian 1992;
Kendall et al 1995, 1996; Lyubimkov 1996, 1998). 
We can extract the following main points:
1.-- The OBN stars show significant He-- and N--excesses. OBN stars
are more frequent among stars above 40 M$_{\odot}$ (Walborn 1988;
Sch\"{o}nberner et al 1988; Herrero et al 1992).
2.-- All fast rotators among the O--stars show some
He--excesses (Herrero et al 1992, 1998, 1999; Lyubimkov 1996).
3.-- Although  rather controversial initially, there seems
to be an increase of the He-- and N--abundances with the relative
age (i.e.\ the fraction  $t/t_{\rm MS}$ 
of the MS lifetime spent) for the early
B--type stars (Lyubimkov 1991, 1996; Gies \& Lambert 1992,
see note added in proof; Denissenkov 1994).
Lyubimkov (1996) suggests a sharp rise from $(He/H)$ = 0.08 -- 0.10
to 0.20 in number for O--stars when $t/t_{\rm MS} \geq 0.5-0.7$, while  for
B--type stars the corresponding value rises  to 0.12 -- 0.14.
As to nitrogen, its abundance is estimated to rise to
about 3 times for a 14 M$_{\odot}$ and to 2 for a 
10 M$_{\odot}$ star.
An increase  of the N--abundance by a factor 2 -- 3 for an
O--star with the average $v_{\rm rot}$ of 200 km/s
is  the order of magnitude typically considered as a constraint 
for recent stellar  models (Heger et al 2000). 
4.-- The boron abundances in five  B--stars on the MS 
have been found to be smaller by
at least a factor 3 or 4 than the cosmic meteoritic
value (Venn et al 1996). The boron depletion occurs in stars
which also show N--excess and this is supporting  the idea that
rotational mixing occurs throughout the star (Fliegner et al
1996).

\textsc{ ABUNDANCES IN SUPERGIANTS}

The main observations  are the following ones: 1.-- He-- and N--excesses
seem to be the rule among OB--supergiants (Walborn 1988). According
to Walborn, only the small group of the ``peculiar'' OBC stars has
the normal cosmic abundances. An excess of He, sometimes
called the ``helium discrepancy'', and corresponding excesses of
N have been found
by a number of authors (Voels et al 1989; Lennon et al 1991;
Gies and Lambert 1992;
Herrero et al 1992, 1999 ; Smith and Howarth 1994;
Venn 1995ab;  Crowther 1997; McEarlan 1998; McEarlan et al 1999).
As shown by these last authors, 
the  determination of the helium abundance 
also depends on the adopted 
value for microturbulence. Villamariz \& Herrero (1999)
and Herrero et al (1999)
point out however that the helium discrepancy is only
reduced, but not solved when microturbulence is accounted for.
2.-- Evidence of highly CNO processed material is present 
for B--supergiants in the range 20 -- 40 M$_{\odot}$, (McEarlan
et al 1999).
Values of  [N/H] (i.e.\ the difference in log with respect to
the solar values) amounting to 0.6 dex have been found for B--supergiants 
around 20 M$_{\odot}$ (Venn 1995ab).
Such values are in agreement with the enrichments found in
the ejecta of SN 1987A (Fransson et al 1989).
3.-- The values of [N/H] for galactic A--type supergiants
around 12 M$_{\odot}$ lie between 0 and 0.4 dex (Venn 1995ab, 1999).
All these values are globally consistent with the above results of
Lyubimkov (1996).
Takeda \& Takada-Hidai (1995) have suggested  that these excesses are 
larger for larger masses, a result in agreement with theory
and also recently confirmed by McEarlan et al (1999).
 For the A--type supergiants in the
SMC, the N/H excesses are much larger spanning a range
[N/H] = 0 to 1.2 dex (Venn et al 1998; Venn 1999).
4.-- Na--excesses have been found in yellow supergiants (Boyarchuk 
\& Lyubimkov 1983) and the overabundances also seem to be larger  for
higher mass stars (Sasselov 1986). Two different explanations
have been 
 proposed, one based on the reaction  $^{22}$Ne(p,$\gamma)^{23}$Na
(Denissenkov 1994), the other one based on 
$^{20}$Ne(p,$\gamma)^{21}$Na (Prantzos et al 1991), with some
additional mixing processes in both cases. The latter reaction
seems to have a too low rate, while the first one may work
(Denissenkov 1994). The important point is that the 
observed Na excesses imply some mixing from the deep interior to the
surface.
5.-- There are only very few abundance determinations in
 yellow and red supergiants.
Some excesses of N with respect to C and O  have been found by 
Luck (1978). Barbuy et al (1996) found both N--enrichments
 and normal compositions among the slow 
rotating  F--G supergiants.  Isotopic
ratios $^{13}$C/$^{12}$C, $^{17}$O/$^{16}$O and $^{18}$O/$^{17}$O
for red supergiants would provide
very useful information. However, the dilution factor in the 
convective envelope of red giants and supergiants is so large
that it is not possible from the rare data available to make any 
conclusion about the presence of additional mixing (Maeder
1987; Dearborn 1992; Denissenkov 1994; El Eid 1994).

\subsection{COMPARISONS OF MODELS AND OBSERVATIONS}

\textsc{MASSIVE STARS IN  MS AND POST--MS PHASES}

Let us first recall that  from the comparison of $t_{\rm mix}$ 
and $t_{\rm H}$, it is clear that  mixing  processes are
more efficient in more massive stars. 
For the intermediate mass stars of the
B-- and A--types, there is no
global mixing currently predicted. Often,
the comparisons with the observed abundance excesses for
O-- and B--type stars are  used to
adjust some efficiency factors in the models 
(Pinsonneault et al 1989;  Weiss et al 1988; Weiss 1994;
Chaboyer et al 1995ab; Heger et al 2000). Although not fully
consistent, these  
 approaches are useful to appreciate the 
importance of the various possible effects. The old
prescriptions of Zahn (1983) were applied by
 Maeder (1987), Langer (1992) and Eryurt et al (1994) and they
led to some surface He-- and N--enrichments.

The prescriptions by Zahn (1992) were applied to
the evolution of a 9 M$_{\odot}$ star (Talon et al 1997).
 They found essentially 
no He--enrichment and a moderate enhancement  (factor $\sim 2$)
of N at the stellar surface,
for an initial velocity of 300 km/s. 
Fig.\ 6 illustrates the changes of the N/H ratios from the 
ZAMS to the red supergiant stage for 20 and 25 M$_{\odot}$ stars
(Meynet \& Maeder 2000).
For non--rotating stars, the surface enrichment in nitrogen
only occurs when the star reaches the red supergiant phase;
there, CNO elements are dredged--up by deep convection.
For rotating stars,  N--excesses  occur already during
the MS phase and they are larger for high rotation and initial
stellar masses. At the end of the MS phase, for solar metallicity
Z = 0.02, the predicted excesses
amount to factors 3 and 4  
for initial $v_{\rm rot}$ = 200 and 300 km/s respectively.
At lower metallicity, the N--enrichment 
during  the MS phase is  smaller,
likely due to the lower mass loss; however, there is 
a very large increase (up to a factor of $\sim 10$)
 for late B--type supergiants, because the star
 spends a lot of time in the blue phase and mixing 
processes have time to work. The predictions of Fig. 6 are in agreement
with the observed excesses for galactic B-- and A--type
supergiants (Venn 1995ab; Venn 1998). Also the very large
excesses observed for A--type supergiants in the SMC
(Venn 1998, 1999) are  remarkably  well accounted for. 

\textsc{QUESTIONS ABOUT NITROGEN }

 Many studies of galactic halo stars,  blue compact
galaxies and  highly redshifted galaxies  have revealed 
the need for initial  production of
primary N in addition to the current secondary nitrogen 
(cf. Edmunds \& Pagel 1978;
Matteucci 1986; Pettini et al 1995; Thuan et al 1995;
 Centurion et al 1998; Pilyugin 1999;
Henry \& Worthey 1999). 
The early production of N in the evolution of galaxies 
seems to imply that some N is produced  
in massive stars (Matteucci 1986; Thuan et al 1995).
The problem is that the usual stellar models  do not show 
such a production without adhoc assumptions.
 Models of rotating stars allow 
us to clearly identify the conditions for the production of primary N: 
the star must have an He--burning core and a thick
and long--lived  H--burning shell, then
diffusion and transport of new $^{12}$C from the core 
to the shell may  generate some primary $^{14}$N. 
W--R stars do not  seem favourable, since the H--shell does not
live long enough, being quickly  extinguished and removed by mass loss.
Low metallicity supergiants that have not  suffered 
large mass loss are a very favourable site (Meynet \& Maeder
2000), especially if rotation is faster at low metallicities. 

\subsection{RED GIANTS AND AGB STARS}

For MS stars in the range of 1.5 to $\sim 10$ M$_{\odot}$, there is no
evidence of extra--mixing, however there are 
interesting indications for red giants.
The study of  $^{12}$C/$^{13}$C in cluster red giants (Gilroy 1989)
shows that stars between 2.2 M$_{\odot}$ and 7 M$_{\odot}$ have
ratios close to the standard predictions without  mixing.
However, red giants below 2.2 M$_{\odot}$ show  $^{12}$C/$^{13}$C
ratios much lower than the predictions, indicating some 
 extra--mixing (Gilroy 1989; see also Harris et al 1988);
the lower the mass, the higher the mixing.
On the red giant  branch of M67, there are indications of
extra--mixing for stars  brighter than 
$\log L/L_{\odot} \simeq 1.0$, where the first dredge--up occurs 
(Gilroy \& Brown 1991). 

From the data on M67, it appears that  extra--mixing is only
efficient when the H--burning shell  reaches  the 
$\mu$--discontinuity left by the inwards progression of the outer
convective zone (Charbonnel 1994). Prior to this stage,
the $\mu$--gradient created by the first dredge--up acts as
a barrier to any mixing  below the convective envelope
(Charbonnel et al 1998). The $\mu$--gradient
necessary to prevent mixing is found to be in agreement with that
expected to stop the meridional circulation. For stars with 
$M \geq 2.2$ M$_{\odot}$, helium ignition occurs in  non--degenerate
cores, i.e.\ early enough so that  the H--shell does not  reach the border
left by the outer convective zone and there is always a 
$\mu$--gradient high enough to prevent mixing (Charbonnel et al 1998).
Boothroyd \& Sackmann (1999) confirm that extra--mixing 
and the associated CNO nuclear processing (cool bottom processing,
CBP) occur when the H--burning shell erases the $\mu$--barrier
established by the first dredge--up  and they predict
that the effects of the CBP behave like 
$M^{-2}$ and $Z^{-1}$. Further observations of red giants with various 
masses are very much needed to confirm the relative absence of
enrichment for masses $\geq  2.2$ M$_{\odot}$.

For the more advanced stages of intermediate mass stars,
the critical questions concern nucleosynthesis and the 
processes leading to the production of the s--elements in  
AGB stars (Iben 1999).
Rotation  appears to allow the formation of larger degenerate cores
(Sackmann \& Weidemann 1972; Maeder 1974), then the 
C/O core mass is further increased during the 
early AGB phases (Dominguez et al 1996). The large 
$\Omega$--gradients between 
the bottom of the convective envelope and the H--burning shell
can drive mixing, mainly by the GSF instability 
and to a lesser extent
by shear and meridional circulation between the
H-- and $^{12}$C--rich layers
during the third dredge--up in AGB stars (Langer et al 1999).
The neutron production by $^{13}$C($\alpha$,n)$^{16}$O
between the thermal pulses is favourable to the production
of s--elements. 
For  further studies on the role of rotation in  AGB stars,
the exact treatment of the instabilities in regions
of steep $\Omega$-- and $\mu$--gradients will play a crucial role.

\section{POST--MS EVOLUTION WITH ROTATION}

The post--MS evolution of rotating stars 
differs from that of non--rotating stars for
three main reasons : 1) the structure at He--ignition 
is different due to 
the rotationally induced mixing during the previous H--burning
phase. In rotating stars, the He--cores
are more massive (Sreenivasan \& Wilson 1985b; 
Sofia et al 1994) and the radiative
envelope is enriched in CNO--burning products (Maeder 1987; 
Heger et al 2000; Meynet \& Maeder 2000). 
2) The mass loss rates are increased by rotation
(Friend \& Abbott 1986; Langer 1998; Sect 2.6).
3) Rotational transport mechanisms may also operate 
in the interior during the post--MS phases. 
However in massive stars,
the timescales  for mixing and circulation, $t_{\rm mix}$
and $t_{\rm circ}$, are much larger than the evolutionary timescale by
one to two orders of magnitudes during the He--burning phase 
(Endal \& Sofia 1978) and even larger in the post
He--burning phases (Heger et al 2000). 
Thus these processes will globally  have small
effects during these stages. However, due to very high angular
velocity gradients occuring locally, some instabilities may appear
on much smaller timescales (Endal \& Sofia 1978; Deupree 1995).

\subsection{INTERNAL EFFECTS}

The fast contraction of the core and  expansion of the 
envelope which follows
the end of the MS phase produces an acceleration of $\Omega$
in the inner regions
and a slowing down in the outer layers
(Kippenhahn et al 1970; Endal \& Sofia 1978; 
Talon et al 1997; Heger et al 2000;
Meynet \& Maeder 2000). Typically
for a  20 M$_\odot$ model with an initial
$v_{\rm rot}=300$ km/s  (Fig. 5), the ratio
of the central to surface
angular velocity never exceeds 5 during the MS phase,
while it increases up to 10$^5$ or 10$^6$ during 
the He--burning phase (Sect.\ 5.5).

At the beginning of the He--burning phase, the He--cores in rotating
models are more massive by about 15\% for $v_{\rm rot}= 300$ km/s. 
The further evolution of the convective core 
 depends on the adopted criterion for convection.
If the Ledoux criterion is used, the growth  of the convective 
core is prevented by the $\mu$--barriers and it 
remains small. Above it several
small convective zones appear, each separated by semiconvective layers
(e.g. Langer 1991c; Heger et al 2000). If the
Schwarzschild criterion is used, the convective core simply grows
in mass. For both cases, the rotational effects
depend on their sensitivity to the 
$\mu$--gradients. As an example, models 
with the Ledoux criterion
show that when rotational mixing is artificially made
 insensitive to $\mu$--gradients, the shear mixing efficiently operates in 
semiconvective regions and  considerably enlarges the final C/O 
core masses (Heger et al 2000).
Typically the C/O core mass increases from a value of 1.77 M$_\odot$ in the
15 M$_\odot$ non--rotating model to a value of 3.4 M$_\odot$ for an initial
$v_{\rm rot}= 200$ km/s.
For a similar rotating model using the Schwarzschild criterion and
incorporating the inhibiting effect of the $\mu$--gradients,
 Meynet \& Maeder (2000) obtain a C/O core mass
of 2.9 M$_\odot$ to be compared with the value of 
2 M$_\odot$ obtained in the non--rotating model.
The treatment of the $\mu$--gradient  is thus critical, since
the C/O core masses play a key  role in determining 
the stellar remnants as well as the chemical yields (Sect. 5.5).

\subsection{EVOLUTION IN THE HR DIAGRAM, LIFETIMES, ROTATIONAL
VELOCITIES}

Rotating as well as non--rotating models with initial masses 
between 9 and 40 M$_\odot$ at solar metallicity
evolve towards the red supergiant (RSG) stage after the 
MS phase (Kippenhahn et al 1970; Sofia et al 1994; Heger 1998;
Meynet \& Maeder 2000). 
Due to the larger He--cores, the  rotating stars have 
higher luminosities, as long as mass loss
is not too large. The initial distribution of 
the rotational velocities will thus introduce 
some scatter in  the luminosities of the supergiants originating
from the same initial mass. The lower 
the sensitivity of the  mixing processes to the 
$\mu$--gradients, the greater the scatter.
For initial $v_{\rm rot}$ between 0 and 300 km/s, the
difference  will be of the order of
0.25 mag (Fig. 5). 

In rotating stars of initial mass between 20 and 40 M$_\odot$, 
due to the large cores, the quantity of nuclear fuel
is larger, but the luminosities are also higher
and thus the He--burning lifetimes change slightly;
as an example for  initial $v_{\rm rot}$ = 200 -- 300 km/s,
the changes are less than 5\%. 
The ratios $t_{\rm He}/t_{\rm H}$ of the He to H--burning lifetimes 
are  not very sensitive to
rotation and they  remain around 10\% (Heger 1998; Meynet \& Maeder 2000).

\textsc{THE NUMBER RATIO OF BLUE TO RED SUPERGIANTS}

The variation with metallicity Z of the number of blue and red 
supergiants (RSG) is important in relation
to the nature of the supernova
 progenitors in different environments (cf.\ Langer 1991bc)
and population synthesis 
(e.g.\ Cervino \& Mas-Hesse 1994; Origlia et al  1999). The
observations show that the number ratio (B/R) of blue to red
supergiants increases steeply with Z. 
Cowley et al  (1979) examined the variation of the B/R ratio
across the Large Magellanic Cloud (LMC) and found that it increases
by a factor 1.8 when the metallicity is larger  by a factor 1.2.
For M$_{\rm bol}$ between -7.5 and -8.5, the B/R ratio is up to 40
or more in inner Galactic regions and only about 4 in the SMC
(Humphreys \& McElroy 1984). A difference in the B/R ratio 
of an order of magnitude between the Galaxy and the 
Small Magellanic Cloud (SMC) was
 also found  from cluster data
(Meylan \& Maeder 1982). Langer \& Maeder (1995) compared 
different stellar models with the  observations and 
concluded that most  massive star models
have problems reproducing this observed trend.

A part of this difficulty certainly arises from the fact 
that supergiants are often close to a neutral state between 
a blue and a red location in the HR diagram. 
Even small changes in mass loss, in convection or other mixing processes 
greatly affect the evolution and the balance between the red and the blue 
locations (Stothers \& Chin 1973, 1975, 1979, 1992ab; Maeder 1981; 
Brunish et al 1986; Maeder \& Meynet 1989;
Arnett 1991; Chin \& Stothers 1991; Langer 1991bc, 1992;
Salasnich et al  1999). As said by Kippenhan \& Weigert (1990),
 ``the present phase
is a sort of magnifying glass, revealing relentlessly the 
faults of calculations of earlier phases.''

The choice of the criterion for convection plays a key role, 
particularly  when the mass loss rates are small. Models 
with the Ledoux criterion, with or
without semiconvection,  predict at low metallicity
(Z between 0.002 -- 0.004) both red and blue supergiants.
However, when the metallicity increases, the B/R ratio decreases
in contradiction with the observed trend (Stothers \& Chin 1992a;
Brocato \& Castellani 1993; Langer \& Maeder 1995).
Models with the Schwarzschild criterion, with or without overshooting,
can more or less reproduce the observed B/R ratio in the solar
neighborhood. However, they predict no
 or very few red supergiants at the metallicity of the SMC, while many are observed (Brunish et al 1986; 
Schaller et al 1992; Bressan et al 1993; Fagotto et al 1994).

When the mass loss rates are low (i.e. at low Z), 
a large intermediate convective zone forms in the vicinity 
of the H--burning shell, homogenizing part of the star 
and maintaining it as a blue supergiant
 (Stothers \& Chin 1979; Maeder 1981).
For larger mass loss rates, the intermediate
convective zone is drastically reduced and the formation of RSG  favoured. 
A further increase of the mass loss rates may
bring the star back to the blue. When the He--core encompasses 
more than some critical mass fraction $q_c$ of the total mass
(Chiosi et al  1978; Maeder 1981), the star moves to 
the blue and  becomes either a blue supergiant or a W--R star
(e.g. Schaller et al 1992; Salasnich et al  1999; Stothers \& Chin 1999).
The critical mass fraction  $q_c$ is equal to $67\%$ at 60 M$_\odot$, 
$77\%$ at 30 M$_\odot$ and 97\% at 15 M$_\odot$ (Maeder 1981).
This agrees with the investigations made for
lower masses by Giannone et al  (1968).

Rotation mainly affects the B/R ratio through its effect on the interior structure
and on the mass loss rates. Maeder (1987), Sofia et al
(1994) and Talon et al (1997) show
that the effect of additional mixing due to rotational 
instabilities in some respect mimics  that of a small amount of
convective overshoot, which does not favour the formation of RSG
at low Z.  
However fast rotation implies also higher mass loss rates by 
stellar winds and in general this favours the  formation of RSG.
In view of these two opposite effects,
it is still uncertain whether rotation may solve the B/R problem. 
Due to the initial distribution of 
rotational velocities, one expects a scatter of the mass loss rates and therefore different
evolutionary scenarios for a given initial mass star 
(Sreenivasan \& Wilson 1985a). 
However, by producing high mass loss rates even at 
low metallicity, rotation may help to resolve the B/R problem. 

Is rotation  responsible for the observed characteristics
 of the blue progenitor of the SN 1987A ?
The presence of the ring structures around SN 1987A
(Burrows et al  1995; Meaburn et al  1995) which likely result from
axisymmetric inhomogeneities in the stellar winds ejected by the
progenitor (Eriguchi et al 1992; LLoyd et al 1995; Martin \& Arnett 1995), 
and the high level
of nitrogen enhancements in the circumstellar material 
(Fransson et al  1989; Panagia et al 1996; Lundqvist \& Fransson 1996)
are features which may be explained at least in part by rotation.
Woosley et al (1998) suggest  that 
any mechanism that reduces the helium core while
simultaneously increasing helium in the envelope would favour
a blue supernova progenitor.

\textsc{EVOLUTION OF THE SURFACE VELOCITIES}

As already stated in Sect.\ 3.2, as the star evolves 
from the blue toward the RSG
stage, the surface velocities quickly decrease 
(Endal \& Sofia 1979; Langer 1998).
For the stars shown in  Fig. 5, velocities between 
20 -- 50 km/s are obtained when $\log T_{\rm eff} = 4.0$.
Values of the order of 1 km/s are reached at the RSG stage.
Observations confirm this rapid  decline of the surface velocities
(Rosendhal 1970; Fukuda 1982). The values at
 $\log T_{\rm eff} = 4.0$ are in good agreement with the recent
determinations of  rotational velocities for galactic A--type 
supergiants by Verdugo et al (1999).

When the star evolves back to the blue from 
the RSG stage, as it is the case for the rotating 
12 M$_\odot$ model shown on Fig. 5, 
the rotational velocity approaches
the break--up velocity (Heger \& Langer 1998; Meynet \& Maeder 2000). 
This behavior results from the stellar contraction 
which concentrate
a large fraction of the angular momentum of the star
 (previously contained in the
extended convective envelope of the RSG) 
in the outer few hundredths of a solar mass.
At the maximum extension of the blue loop, the equatorial velocity
at the surface of the 12 M$_\odot$  star (Fig. 5) reaches 
values as high as 150 km/s.
As  the star evolves back towards the RSG stage, the surface velocity
declines again to about  2 km/s. When the star 
crosses the Cepheid instability strip
its surface velocity is between 10 and 20 km/s, well inside
the observed range (Kraft et al 1959; Kraft 1966; Schmidt-Kaler 1982).  

The increase of the surface velocity 
occurs every time a star leaves the Hayashi 
line to hotter zones of the HRD (Heger \& Langer 1998).
In most cases, there 
are observational evidences for
axisymmetric circumstellar matter: disks around T Tauri stars 
(e.g. Guilloteau \& Dutrey 1998),
 bi--polar planetary nebulae (e.g. Garcia-Segura et al  1999),
structures around SN 1987A (e.g. Meaburn et al  1995) and 
rings around W--R stars (e.g. Marston 1997). 

\subsection{THE $\Omega$, $\Gamma$ LIMITS AND THE LBV's}

Over recent years, the problem of the very luminous stars 
close to the Eddington limit and 
reaching the break--up limit (Langer 1997, 1998)
 has been discussed in relation to
the Luminous Blue Variables (LBV) and their origin (see 
Davidson et al 1989; de  Jager \& Nieuwenhuijzen 1992;
Nota \& Lamers 1997). The LBV, also called the Hubble--Sandage
Variables and S Dor Variables,  are extreme  OB supergiants with
$\log L/L_{\rm \odot} \simeq 6.0$ and $T_{\rm eff}$ between
about 10 000 and 30 000 K (Humphreys 1989). Only a few of them
are known in the Milky Way, among which $\eta$ Carinae (Davidson \& Humphreys 1997;
Davidson et al 1997).
They experience giant 
outbursts with shell ejections. Often they show surrounding 
 bi--polar nebulae  (Nota et al 1997). Many models and types of instabilities
have been proposed to explain the LBV outbursts (de Jager \&
Nieuwenhuijzen 1992; Stothers \& Chin 1993, 1996, 1997; 
Nota \& Lamers 1997).
 
\textsc{PHYSICS OF THE BREAK--UP LIMIT}

The first problem concerns the expression of 
the break--up limit for stars close to the
Eddington limit, i.e. for the brightest supergiants.
 When the radiation
field is strong, the radiative  acceleration
$\vec{g_{\rm rad}}$ must be
accounted for in the total acceleration
\begin{equation} 
\vec{g_{\rm tot}} =
\vec{g_{\rm grav}} + \vec{g_{\rm rot}} +
\vec{g_{\rm rad}} =
\vec{g_{\rm eff}} + \vec{g_{\rm rad}} \;, 
\end{equation}%eqn.\14
\noindent whith  a modulus 
$g_{\rm rad} = (\kappa L/ 4\pi c R^2)$, $R$ being the
equatorial radius. Thus, the break--up velocity obtained
when $\vec{g_{\rm tot}} = 0$ is found to be
 $v_{\rm crit}^2 = \frac{GM}{R}(1-\Gamma)$ 
(Langer 1997, 1998 ; Langer \& Heger 1998; Lamers 1997),
 where $\Gamma$ is the Eddington factor 
$\Gamma= \kappa L/(4\pi c G M)$. For the most luminous stars,
$\Gamma \rightarrow 1$ and thus the critical velocity
tends towards  zero. This has led Langer (1997, 1998) to conclude that 
for any initial rotation, the critical limit  is reached 
before the Eddington limit. Therefore, Langer claims that one 
should rather speak of an $\Omega$--limit for LBV stars rather
than of a $\Gamma$--limit.

Glatzel (1998) has suggested that the $\Omega$--limit is an artifact
due to the absence of von Zeipel's relation in the expression of
$g_{\rm rad}$. Indeed, with von Zeipel's relation the radiative  flux 
tends towards  zero  when  the resulting gravity is zero. 
Thus the critical velocity is  just  
$v_{\rm crit}^2 = \frac{GM}{R}$, while rotation reduces 
(up to 40 \%; Glatzel 1998) the limiting  luminosity.
Stothers (1999) also considers that fast rotation 
reduces  the limiting luminosity. 

For stars close to the Eddington limit, convection may develop in
the outer layers (Langer 1997; Glatzel \& Kiriakidis 1998).
This is however not an objection to the
application of the  von Zeipel theorem, since most of the flux 
is carried by radiation at the surface. Another  possible objection
(Langer et al 1999) is that according to a generalization of the
von Zeipel theorem by Kippenhahn (1977), the radiative flux at the
equator may be reduced or increased depending on the internal 
rotation law. However, the deviations
from von Zeipel's theorem are negligible in 
the current cases of  models with shellular rotation (Maeder 1999a). 
Thus, a study of the physical conditions, of the critical velocity 
and of the instabilities in rotating stars close to the Eddington limit 
is still very needed.

\textsc{ THE EVOLUTION OF LBV's}

According to the evolutionary models at high masses (Schaller
et al 1992; Stothers \& Chin 1996; Salasnich et al 1999), there are
three possible ways for very massive stars to reach the
$\Omega$--limit in the HR diagram.
1.--  Stars with very large initial mass 
and high rotation, especially if their $v_{\rm rot}$
is increased by blueward evolution during the MS phase,
may reach the  $\Omega$--limit in the blue part 
of the HR diagram. 
Some fast rotators may  reach the break--up limit during the
overall contraction phase at the end of the MS, as shown for the
60 M$_{\odot}$ (Fig.\ 4). If the mass 
loss for O--stars is mainly bipolar (Maeder 1999a), 
the reduction of $v_{\rm rot}$ during the MS phase may be smaller. 
For smaller mass loss rates as in lower metallicity galaxies,
$v_{\rm crit}$ could   possibly be reached  earlier
in evolution. The star $\eta$ Carinae shows evidence that the
$\Omega$--limit  is reached in the blue and is likely
at the end of its MS phase or beyond in view of its surface composition 
(Davidson et al 1986; Viotti et al 1989).
2.-- After the end of the  MS phase, when the
star evolves redwards in the HR diagram,
the value of $\frac{\Omega}{\Omega_{\rm crit}}$ 
becomes quite small, since the star evolves with
essentially local conservation of angular momentum. Thus,
rotation is less important. However, the  $\Gamma$--limit 
without rotation lies at a much lower luminosity there (Lamers 
1997; Ulmer \& Fitzpatrick 1998),
so that the $\Omega$--limit may be reached
by the very massive stars during their redwards crossing
of the HR diagram. 3.-- When   stars leave the red supergiant phase, 
either on blue loops or evolving towards the W--R stage, 
the ratio $\frac{\Omega}{\Omega_{\rm crit}}$ increases
quite a lot. This is due to conservation of 
angular momentum in retreating convective envelopes,
which  contributes to  strongly spin up the bluewards evolving stars
(Langer 1998). Thus, the $\Omega$--limit
may also be reached from the red side.
 This  possibility is particularly interesting since 
the observed CNO abundances in some nebulae around LBV stars
are  the same as in red supergiants, which suggests 
that some LBV may originate from red supergiants (Smith 1997). A similar
conclusion was obtained by Waters (1997), who found evidence in some LBV
nebulae  of crystalline forms of silicates, with composition
similar to that of red supergiants.

\textsc{THE NEBULAE AROUND LBV's: SIGNATURE OF ROTATION ?}

Almost all nebulae around LBV stars show a bipolar structure
(Nota et al 1995; Nota \& Clampin 1997). 
This might be related to binarity
 (Damineli 1996; Damineli et al 1997).
Most models invoke collisions of winds of different 
velocities and densities, emitted at different phases of
their evolution. In some cases, an equatorial density enhancement 
is assumed before the outburst (Frank et al 1995; 
see also Nota et al 1995),
while other models assume rather arbitrary non--spherical winds
or a ring--like structure 
interacting with a previous spherical wind 
(Dwarkadas \& Balik 1998; Frank et al 1998).
The models by Garcia--Segura et al (1996, 1997) 
and Langer et al (1999) consider three
phases in the formation of the nebula for $\eta$ Carinae.
In both the pre-- and post--outburst phases,  the star has the 
spherical   fast and low density wind
typical of a blue supergiant. At the break--up limit, the star is 
assumed to have a slow dense wind
concentrated in the equatorial plane. The bi--polar
structure then arises because the shell ejected in the third phase
expands more easily into the lower density at the pole.
 Langer et al
(1999) assume that the equatorial enhancement during the outburst
results from the wind compressed disk model (Bjorkman \&
Cassinelli 1993),  which does not apply  if the von Zeipel
theorem is used (Owocki \& Gayley 1997, 1998). Nevertheless,
we note that  due  to the ``$\kappa$--effect'' 
at the break--up limit (Maeder 1999a),  a strong equatorial ejection
occurs quite naturally, characterized by a high density and a
low velocity, as required by the above  colliding  wind model of
Langer et al (1999).

\subsection{ROTATION AND W--R STAR FORMATION}

\textsc{GENERALITIES}

Recent reviews on the Wolf--Rayet (W--R) phenomenon have been 
presented by Abbott \& Conti (1987), van der Hucht (1992), 
Maeder \& Conti (1994) and  Willis (1999).
Wolf--Rayet  stars are bare cores of initially massive stars 
(Lamers et al  1991). Their original H--rich envelope has been removed 
by stellar winds or through a Roche lobe overflow in a close binary system.
Observationally, most W--R stars appear to originate from stars initially
more massive than about 40 M$_\odot$ (Conti et al  1983; Conti 1984;
Humphreys et al 1985; Tutukov \& Yungelson 1985), however a few 
stars may originate from initial
masses as low as 15--25 M$_\odot$ (Th\'e et al  1982; 
Schild \& Maeder 1984; Hamann et al  1993; Hamann \& Koesterke 1998a; 
Massey \& Johnson 1998). 
The stars enter the W--R phase as WN stars, i.e.\ with surface
abundances representative of equilibrium CNO
processed material. If the peeling off
proceeds deep enough the star may enter the WC phase,
during which the He--burning products appear at the surface.

Many observed features are well reproduced by current stellar models.
Typically  good agreement is obtained between the
observed and predicted values for the surface abundances of WN stars
(Crowther et al  1995; Hamann \& Koesterke 1998a).
This  indicates the general 
correctness of our understanding of the CNO cycle and of 
the relevant nuclear data (Maeder 1983), but is not a test of the
model structure.
For WC stars,  
comparisons with observed surface abundances also generally 
show a good agreement (Willis 1991; Maeder
\& Meynet 1994). In particular, the strong surface Ne--enrichments 
predicted by the models of WC stars have been confirmed 
by ISO observations (Willis et al 1997, 1998; Morris et al 1999; Dessart et al 1999). 

The star number ratios W--R/O, W--R/RSG, WN/WC show a
strong correlation with metallicity 
(Azzopardi et al  1988; Smith 1988; 
Maeder 1991; Maeder \& Meynet 1994; Massey \& Jonhson 1998).
For instance, the W--R/O number ratio increases
with the metallicity Z of the parent galaxy.
Despite many other
claims (Bertelli \& Chiosi 1981, 1982; Garmany et al 1982; 
Armandroff \& Massey 1985; Massey 1985; Massey et al 1986),
the main cause is metallicity Z, which 
through stellar winds influences stellar evolution
 and thus the W--R lifetimes (Smith 1973; 
Maeder et al  1980; Moffat \& Shara 1983).
The higher the metallicity, the stronger is the mass loss 
by stellar winds and thus the earlier is the entry in
the W--R phase for a given star; also 
the minimum initial mass for forming a W--R star
is lowered. 

\textsc{REMAINING PROBLEMS WITH W--R STARS}

Despite these successes, observations  indicate
some remaining problems:
1.-- It is possible to reproduce the
W--R/O and  WN/WC number
ratios observed in the Milky Way and in various galaxies
of the Local Group, only by using models with 
 mass loss rates enhanced by a factor of two  during the MS
and  WNL phases (Maeder \& Meynet 1994).
The relative populations of WN and WC stars observed 
in young starburst regions are also better
reproduced when models with high mass loss rates are 
used (Meynet 1995; Schaerer et al  1999).
This is not satisfactory, since  clumping 
in the winds of hot star tends to reduce by a factor 2 -- 3
the  observed mass loss rates (Nugis et al  1998;
Hamann \& Koesterke 1998b).
2.-- The lower limit for the luminosities of WN stars 
(around $\log L/L_\odot \sim
5.0$;  Hamann \& Koesterke 1998a) is fainter 
than predicted by standard evolutionary tracks.
Massey \& Johnson (1998) find that the presence of luminous 
red supergiants (RSG) and W--R stars is 
well correlated for the OB associations in M31 and M33, suggesting that 
some stars with mass $\geq$ 15 M$_\odot$  go through both the 
RSG and W--R phases.
3.-- For WN stars, there is a continuous transition from high
H--surface abundances (0.4--0.5 in mass fraction) to hydrogen--free
atmospheres, while standard models predict an abrupt transition
(Langer et al  1994; Hamann \& Koesterke 1998a; Fig. 7, Meynet 
\& Maeder 2000).
4.-- Smith \& Maeder (1998) show that, besides the mass,
 a second parameter affecting
the mass loss rates and terminal velocities  of the wind is necessary
to characterize the hydrogen--free WN stars.
5.-- Standard models
do not reproduce the observed number of 
stars in the transition WN/WC phase, characterized by
spectra with  both H-- and He--burning products. 
These models predict indeed an abrupt
transition from WN to WC stars, since the He--core is growing 
and thus building up a steep chemical discontinuity at its
outer edge (e.g. Schaller et al  1992). Thus,  
almost no
($<$ 1\%) stars with intermediate characteristics of WN
and WC stars are predicted. However, 4 -- 5\% of the W--R stars are in such 
a transition phase (Conti \& Massey 1989; van der Hucht 1999),
showing  that some extra--mixing  is at work (Langer 1991b).

\textsc{ROTATION AND THE FORMATION OF W--R STARS}

Rotation  may  affect
the formation and properties of W--R stars
in several  ways (Sreenivasan \& Wilson 1982, 1985a; 
Maeder 1987; Fliegner \& Langer 1995; Maeder 1999b; Meynet 1999):

\noindent 1.-- 
Surface abundances characteristic
of the WNL stars may appear in a rotating star, not only as a result of 
the mass loss which uncovers inner layers, but also as a result 
of  mixing in radiative zones.  
The same remark applies to  the entry into the WC phase. 

\noindent 2.-- Rotation may imply different evolutionary scenarios.
Before becoming a W--R star, the non--rotating 
60 M$_\odot$ model at solar metallicity
is likely to go  through a short
LBV phase after the H--exhaustion in its core. 
In the case of fast rotation, the star may
enter  the W--R phase, still burning hydrogen in its core
(Maeder 1987; Fliegner \& Langer 1995; Meynet 1999),
thus skipping  the LBV phase and spending more time in the W--R phase.

\noindent 3.-- Rotation favours the formation of  W--R stars 
from lower initial mass both
through its  effects on the mass loss rates 
(Sreenivasan \& Wilson 1982; Sect. 2.6) and on the mixing.
Typically for the non--rotating models shown in Fig.\ 5,
the minimum mass for W--R star formation is betwen 35--40 M$_\odot$.
It decreases to about 25 M$_\odot$ for initial $v_{\rm rot}=300$ km/s.
This effect may help  to explain the low luminous WN stars 
reported by Hamann \& Koesterke (1998a). It also favours the
entry into the W--R phase from the  RSG stage.

\noindent 4.-- During the WN phase, the surface abundances are different. 
Indeed as a
consequence of the first point, the N/C, N/O ratios
obtained at the surface of the rotating WN models
may have not yet reached 
the full nuclear equilibrium in contrast
with the non--rotating case where nuclear equilibrium is 
reached as soon as the star enters the WN phase. 
The CNO ratios  are however close to the equilibrium values
(Fig. 7). During the transition WN/WC phase,
nitrogen enhancements can be observed 
simultaneously with carbon and neon enhancements. 
After this transition phase, the $^{22}$Ne enhancement 
reaches more or less the same high equilibrium level
whatever the initial angular velocity, in agreement 
with the determinations
of the neon abundance at the surface of WC stars (Willis et al  1997).

\noindent 5.-- 
Higher  rotational velocities lead to longer W--R lifetimes.
As an example for a 60 M$_\odot$ model (Fig. 7),
the W--R lifetime is increased by more than a factor 3
when rotation is included.
The durations of the WN and of the transition WN/WC phases
are increased. The ratio of the   
lifetimes of the WC to the WN
phase is reduced. 

\noindent 6.-- 
High rotations lead to  less luminous WC stars.
 This is because a fast rotating star
enters the W--R stage  earlier in its evolution and thus begins 
to loose large amounts of mass early. Therefore, fast rotators
enter the  WC phase with a small mass and a low luminosity;
the final masses are also smaller.

Thus rotation could remove or at least
alleviate the  above mentioned problems.
The need to enhance the mass loss rates to reproduce the
observed W--R/O number ratios  no longer appears necessary.
Rotation also implies effects which cannot be reproduced by 
an increase of the mass loss rate. In particular,
mixing induced by rotation produces
milder chemical gradients
and leads to a more progressive decrease of the
hydrogen abundance at the surface of WN stars (Fig. 7). 

\textsc{ARE W--R STARS FAST OR SLOW ROTATORS ?}

Direct attempts to measure the rotational velocity of W--R stars have been 
performed only for a few cases: Massey (1980) and
 Koenigsberger (1990) obtain $v \sin i \sim 500$ km/s for WR138. 
However the binary nature of this
 object (Annuk 1991) blurs this picture, since the origin of this
 high velocity might be the O--type companion.
The second case, WR3 with  $v \sin i \sim$ 150 -- 200 km/s,
looks more promising (Massey \& Conti 1981) since the broadened 
absorption lines move in phase with the W--R emission lines
(Moffat et al  1986).

There is some indirect evidence pointing towards the existence
of some axisymmetric features around W--R stars (see Drissen 1992;
 Marchenko 1994). For instance Arnal (1992) has mapped
the environment of 6 W--R stars at 
a frequency of 1.42 MHz and found that all the HI cavities 
around these have an elongated shape with a mean
major--to--minor axis ratio of about 2.2.
Other evidences have been found by Schulte--Ladbeck et al  (1992), 
Miller \& Chu (1993). According to Harries et al  (1998), 
about 15\% of W--R stars have anisotropic winds.
They suggest that the main cause of the wind anisotropy 
is equatorial density enhancements produced by fast
rotation rates and  estimate the rotational velocities to be 
about 10 -- 20\% of the break--up velocity. 

The surface velocities of W--R stars depend mainly on the initial velocity 
and on the amount of angular momentum lost during the previous stages.
This amount will depend  on the exact evolutionary sequence followed;
in particular, the questions are,  whether the star has passed 
through the RSG stage and what were the anisotropies of the 
stellar winds. Some  other effects may also 
intervene, for instance  the  possible presence of a magnetic 
field (e.g. Cassinelli 1992; Sreenivasan \& Wilson 1982; Sect. 3.2).

For the 60 M$_\odot$ model
shown on Fig. 7 with $v_{\rm rot}$ = 300 km/s,
computed assuming spherically symmetric winds and no magnetic fields, 
the surface velocity ranges between 20 and 40 km/s,
i.e. between 3 and 6\% of the break--up velocity
during most of the WNL phase. At the
beginning of the core He--burning phase, the He--core contracts, the 
small H--rich envelope expands and the
surface velocity reaches the 
break--up limit.
Huge mass loss rates ensue, which eject about 3 M$_\odot$ of
material forming an anisotropic nebula with
abundances characteristic of CNO--equilibrium. 
When the star has lost sufficient angular momentum,
it drops below the break--up limit and pursues its evolution with a
nearly constant surface
velocity around 40 km/s. The contraction of the W--R star at the
very  end of the He--burning phase may again increase
 $v_{\rm rot}$, but it remains far below the break--up limit 
($\Omega/\Omega_{\rm crit} \sim$ 5\%).

\subsection{LATE STAGES, REMNANTS AND CHEMICAL YIELDS}

\textsc{THE POST He--BURNING PHASES}

The masses of the C/O cores are larger in rotating stars, that do 
not evolve through a W--R phase (Sofia et al 1994; Heger et al 2000).
As an example, at the end of the He--burning phase, the C/O core mass 
in a rotating 20 M$_\odot$ model with an initial 
$v_{\rm rot}=$ 300 km/s is 5.7 M$_\odot$ (Meynet \& Maeder 2000). 
The value in a  non--rotating model is 3.8 M$_\odot$. 
Thus, a rotating 20 M$_\odot$ star will have
a behavior during the late stages similar to that of 
a non--rotating 25 M$_\odot$ star. 

It is also interesting to notice that
due to the larger He--cores, the $^{12}$C($\alpha$,$\gamma$)$^{16}$O 
reaction is more active at the end of the He--burning phase;
therefore the fraction of carbon
left in the C/O core decreases with respect to that in
the non--rotating model (by about
a factor 2.5 in the  above example). This leads to an
increase in the oxygen yield. Moreover since the carbon burning 
phase is considerably reduced, the stellar 
core has less time to  remove  its entropy through
heavy neutrino losses, favouring the formation of black holes
(Woosley 1986). According to Fryer (1999) the lower mass limit 
for black hole formation is likely lowered by rotation 
(see also Fryer \& Heger 1999).

After central He--exhaustion, the He--shell ignites and 
the above layers expand leading to a decrease of  the strength 
of the H--burning shell. The smaller thermal gradient near the 
H--shell favours the mixing of chemical elements there 
(Heger et al 2000). Protons
as well as nitrogen are brought  from the H--shell down into 
the  underlaying He--rich layers. 
At the same time, due to the contraction of the C/O core, 
the temperature at the bottom of
the He--shell increases and the overlying convective zone extends 
in mass, engulfing the region rotationally enriched in protons 
and nitrogen. These species 
then burn very rapidly, on a  timescale shorter  than
the convective turn--over time; thus, 
they have nearly completely disappeared before  reaching 
the bottom of the He--convective zone (Heger 1998).
The large  extension of the He--burning shell has important
consequences for nucleosynthesis. This extension
is larger for  larger core masses, i.e. for large  initial mass 
and/or for large rotational velocities. 

When evolution proceeds further, the rotating core speeds up more and more,
possibly becoming unstable with respect to nonaxial symmetric
perturbations (Kippenhahn et al 1970; Ostriker \& Bodenheimer 1973;
Tassoul 1978). However, the results obtained by Heger
et al (2000) for stars with initial $v_{\rm rot}$ between 200 and 300 km/s 
suggest that, before the core collapse, the ratio of 
the rotational to the  potential energy is lower than required for
such instabilities to occur. 

\textsc{ROTATIONAL PERIODS OF PULSARS}

According to the
models by Heger et al  (2000) and Meynet \& Maeder (2000),
at their birth neutron stars (NS) 
should have rotation periods of about 0.6 ms,
being nearly at the break--up rate. What is striking 
is that these periods are much smaller than the
measured periods for young pulsars, which are around 
20 -- 150 ms (Marshall et al 1998). This means that the 
models have between $\sim$ 20 and 100 times more specific
angular momentum than found in the young neutron stars. 
Various effects may be responsible for this excess rotation
in the models. The efficiency of some rotationally induced mixing 
processes may have  been underestimated
or some important transport mechanism may still be missing. 
Kippenhahn et al (1970) speculated about the possibility for
rapidly rotating dense cores to shed some mass into the
envelope at its equator, in a way 
similar to rapidly rotating stars shedding  mass into 
the circumstellar envelope. The equatorial mass loss by
anisotropic stellar winds  heavily modifies the
surface boundary conditions and may remove a huge amount of
angular momentum (Maeder 1999a).
Other braking mechanisms, such as the removal of angular momentum  
from the convective core by gravity waves (Denissenkov et al 1999; 
Sect. 2.5) or through a magnetic field (Spruit \& Phinney 1998)
may also be invoked.
As  pointed out by Fricke \& Kippenhahn (1972),
the coupling of the core and envelope  cannot be complete,
because with solid body rotation at all time, the core would
rotate too slowly (P $\simeq$  650 ms) to form pulsars with the
observed periods. 

The evacuation of the excess  angular momentum could also
have occured during the formation of a NS. The NS
could  also have been born spinning at break--up velocity and have
been very efficiently slowed down during the first years. 
However, as discussed by Hardorp (1974),
there are arguments suggesting that NS have never been near
break--up (Ruderman 1972), and the study of the  Crab pulsar supports 
this view (Trimble \& Rees 1970). Indeed in this case, the
release of such an important amount of rotational energy,
if not emitted in the form of a $\gamma$--ray burst or through
gravitational waves,
would have shown up in the expansion energy of the nebula and in 
the optical light during historical times, which is not reported.
The original rotation period of the Crab pulsar at birth is
estimated to be  5 ms, still one order of magnitude from the 
break--up period (Hardorp 1974). Therefore, the stellar core must 
have been spinning slowly before its collapse. 

If such a large angular momentum is embarrassing at present for 
explaining the observed rotating periods of young pulsars, it
may give some support to the  ``collapsar'' model 
proposed by Woosley (1993; MacFadyen \& Woosley 1999) for
the $\gamma$--ray bursts. 
A ``collapsar''
is a black hole formed by the incomplete explosion of a rapidly rotating
massive star. The rapid rotation is necessary to allow the formation
of an accretion disk outside the black hole. The accretion disk 
efficiently transforms the 
gravitational binding energy into heat which can then power
a highly relativistic jet. The burst and its afterglow in 
various wavelengths is attributed to the jet
and its interactions with the external medium.
The models by Heger et al  (2000) and Meynet \& Maeder (2000)
have enough angular momentum to support matter in a stable disk
outside a black hole and thus could offer
interesting progenitors for this kind of evolution, if it exists.

\textsc{THE CHEMICAL YIELDS}

Rotation affects the chemical yields in many ways. 
The larger He--cores obtained in rotating models at core collapse 
imply larger production of helium and other $\alpha$-nuclei elements
(Heger 1998). This is by far the most important effect of
rotation on the chemical yields.
Also, by  enhancing the mass loss rates and by 
making the formation of W--R stars easier, rotation favours  the
enrichment of the interstellar medium by stellar winds
(Maeder 1992). Indeed, the stronger 
the winds, the richer the ejecta in helium and carbon and the lower
in oxygen.

The rotational diffusion  during the H--burning phase enriches 
the outer layers in CNO processed elements (Maeder 1987;
Fliegner \& Langer 1995; Meynet 1998; Heger 1998). 
Some $^{14}$N  is extracted from the core and saved from 
further destruction. The same can be said
for $^{17}$O and $^{26}$Al, a radioisotope with a half--life of 0.72 Myr. 
The mixing in the envelope of rotating stars also leads  
to a faster depletion of the temperature sensistive
light isotopes, for instance lithium and boron (Fliegner et al 1996).

The presence of $^{26}$Al in
the interstellar medium is responsible for the diffuse galactic 
emission observed at 1.8 MeV (e.g. Oberlack et al  1996).  
If the nucleosynthetic sites of this element
appear to be the  massive stars (Prantzos \& Diehl 1996), 
it is still not yet clear how the production is shared between 
the supernovae and the W--R stars and how it is affected by 
rotation and binarity.
For stellar masses between 12 and 15 M$_\odot$, the lifetimes are 
much longer than that of $^{26}$Al and therefore most of the
$^{26}$Al produced during central H--burning and partially
mixed in the envelope has decayed at the time of the supernova 
explosion. Thus, for this mass range, rotation does not seem to 
bring important changes (Heger 1998). However, when the star is massive
 enough to go through the  W--R phase, the stellar winds
may remove $^{26}$Al--enriched layers at a much earlier stage.
In that case, rotation may substantially increase
the quantity of $^{26}$Al injected in the interstellar medium
(Langer et al  1995).

The convective zone associated with the He--shell in rotating models
transports H--burning products to the He--burning shell (Heger 1998).
The injection of protons and nitrogen into a He--burning zone opens 
new channels of nucleosynthesis (Jorissen \& Arnould 1989). In
particular, this enhances the  s--process and the formation 
of $^{14}$C, $^{18}$O and $^{19}$F. 
Since these elements are produced just before the core collapse 
they can survive until the supernova explosion. The injection 
of protons into a He--burning 
zone may also be responsible for primary $^{14}$N
production through $^{12}$C(p,$\gamma)^{13}$N$(\beta^+) ^{13}$C(p,$\gamma)^{14}$N, 
but this 
primary nitrogen is rapidly destroyed to produce
$^{18}$O. As discussed in Sect. 4.2, stars  
with an He--burning core and a thick and long--lived H--burning shell
seem to be more favourable sites for a primary $^{14}$N production.
Another important effect of the 
growth of the He--shell in these late stages is the
production shortly before core--collapse of some $^{15}$N.
In non--rotating models, this element is destroyed, while
in rotating ones it is synthesized (Heger 1998). The very
low $^{14}$N/$^{15}$N ratios measured in  
star--forming regions of the LMC and in the core of the (post--)starburst 
galaxy NGC 4945 (Chin et al 1999) support 
an origin of $^{15}$N in  massive stars. 

\section{PERSPECTIVES}
We hope to have shown that rotation is unavoidable for a 
proper understanding and modelling of the evolution for the
Upper MS stars.

The way to further progress goes through more studies
of the physical effects of rotation, in particular of the 
various instabilities which can produce mixing of the
elements and transport of angular momentum, both in the
early and advanced phases of evolution. The influence of 
rotation on mass loss is also critical. In this respect,
the shape and composition of the asymmetric nebulae observed
around many massive stars provide interesting constraints on the
models of rotating stars.\\
ACKNOWLEDGEMENTS: We express our best thanks to 

Dr. Laura Fullton for her most useful advices on the manuscript.

\begin{figure}
\epsfxsize=12cm  \epsfbox{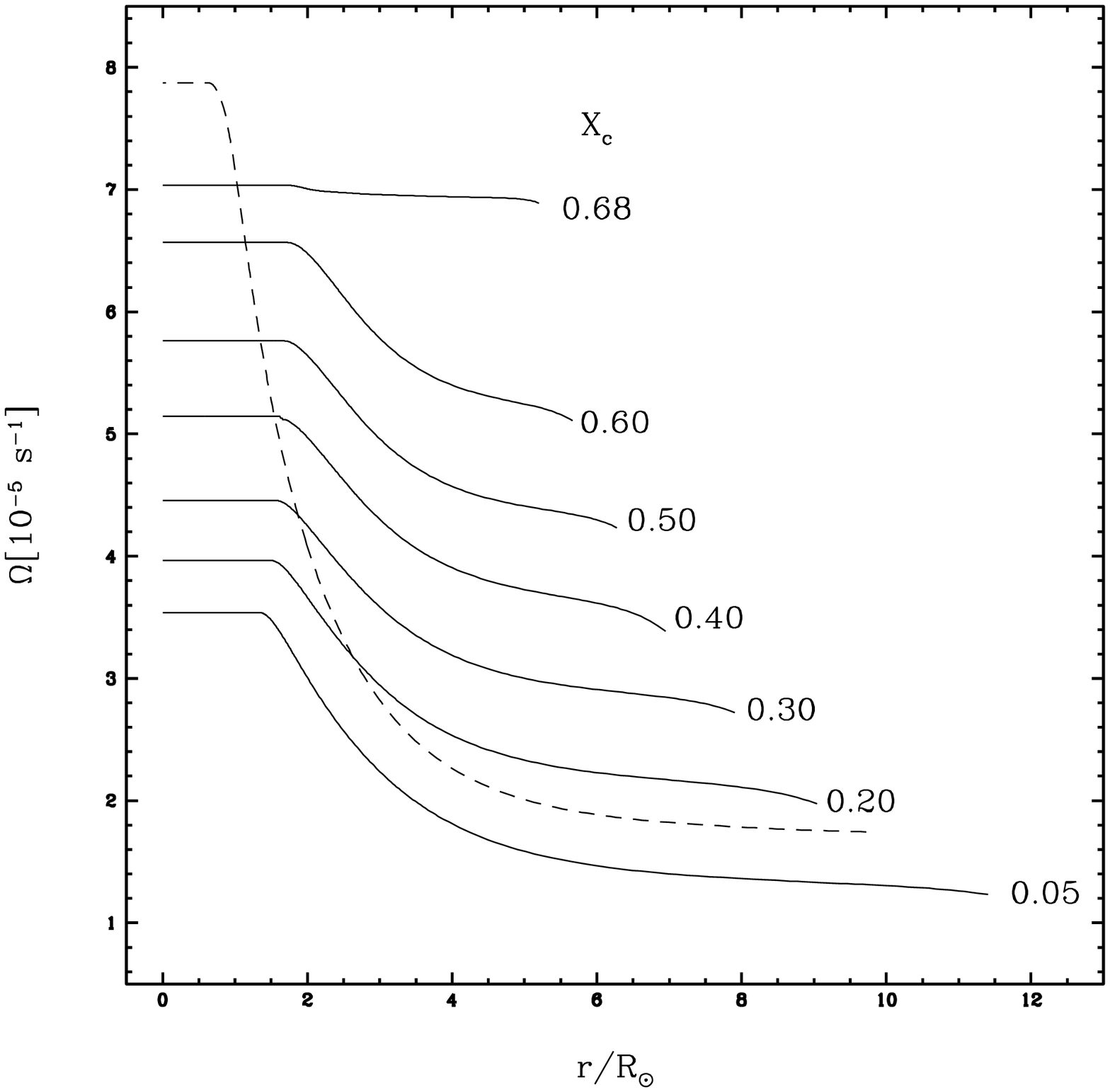}
\caption{Evolution of the angular velocity $\Omega$ 
 as a function of the distance to the center
in a 20 M$_\odot$ star with an initial $v_{\rm rot}$ = 300 km/s. 
$X_c$ is the hydrogen mass fraction at the center.
The broken line shows the profile when the He-core contracts at the end
of the H-burning phase.\label{Fig:AngularVel}}
\end{figure}

\begin{figure}
\epsfxsize=12cm  \epsfbox{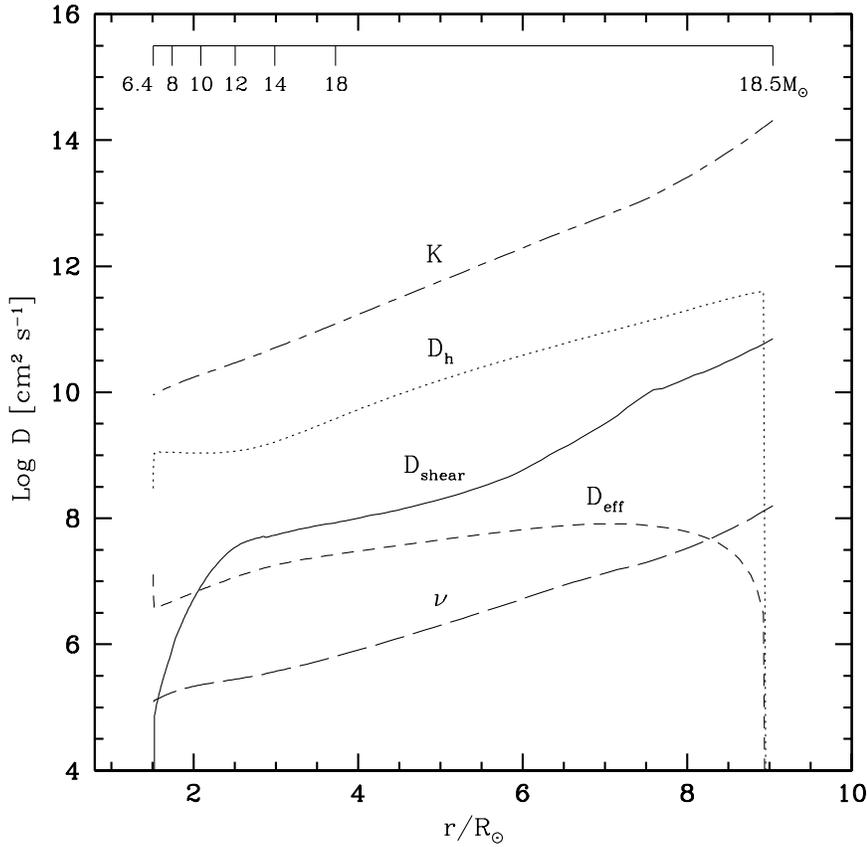}
\caption{Internal values  of $K$ the thermal diffusivity,
 $D_{\rm h}$ the coefficient of
horizontal turbulence, $D_{\rm shear}$ the shear diffusion coefficient,
$D_{\rm eff}$ the effective diffusivity (see text) and $\nu$ the total
viscosity (radiative + molecular)
in the radiative envelope of a 20 M$_\odot$ star 
with an initial $v_{\rm rot}$ = 300 km/s. 
The lagrangian mass coordinate is given on the upper scale.
Here, the hydrogen mass fraction at the center $X_c = 0.20$.}
\end{figure}

\begin{figure}
\epsfxsize=12cm  \epsfbox{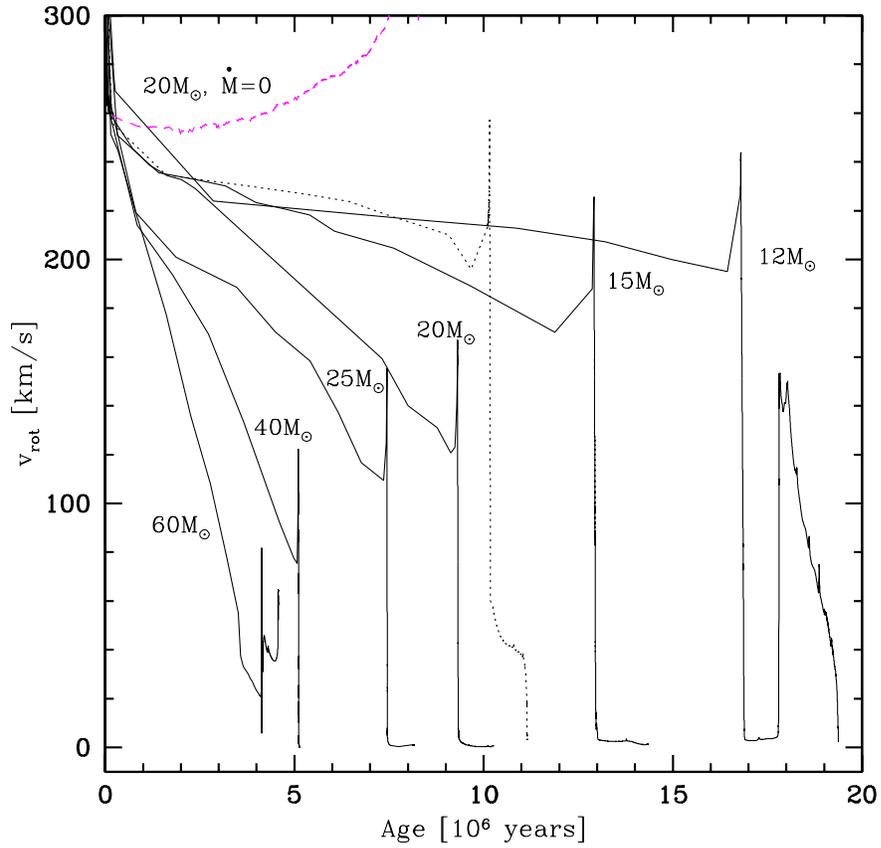}
\caption{Evolution of the surface equatorial velocity as 
a function of time for stars of various initial stellar masses
(Meynet \& Maeder 1999). 
All  models have an initial velocity of 300 km/s.
The continuous lines refer to solar metallicity models, 
the dotted line corresponds
to a 20 M$_\odot$ star with Z = 0.004.   
The dashed line  corresponds to a 20 M$_\odot$ star
 without mass loss.}
\end{figure}

\begin{figure}
\epsfxsize=12cm  \epsfbox{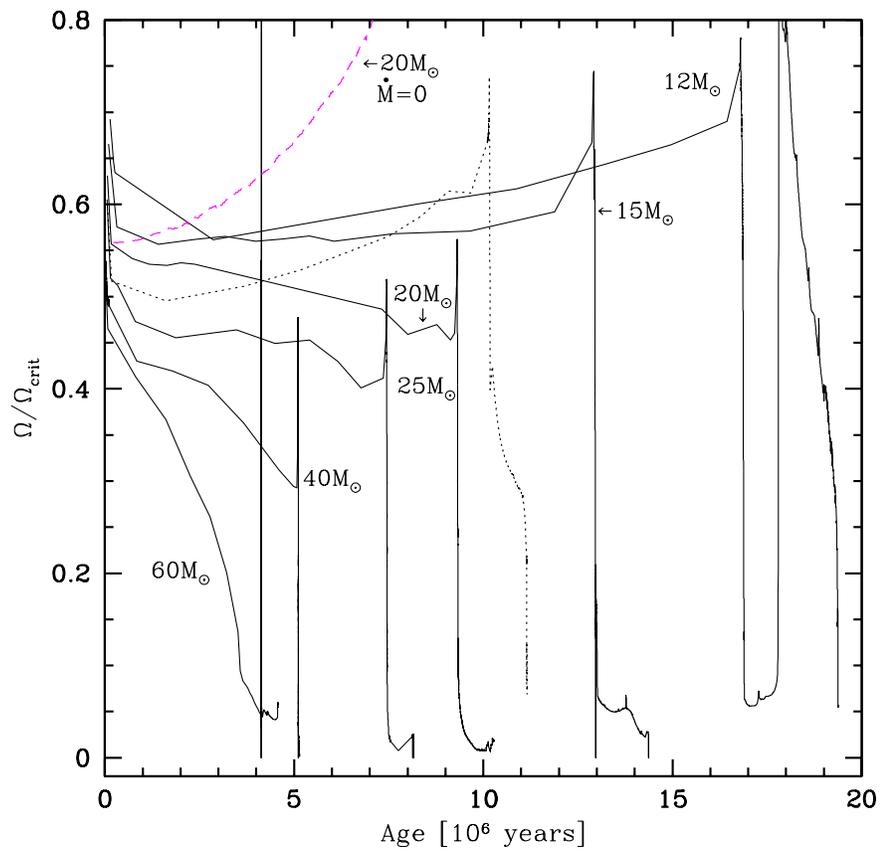}
\caption{Same as Fig. 3 for the ratio $\Omega/\Omega_{\rm crit}$ 
of the angular velocity to the break-up velocity
at the stellar surface.}
\end{figure}

\begin{figure}
\epsfxsize=12cm  \epsfbox{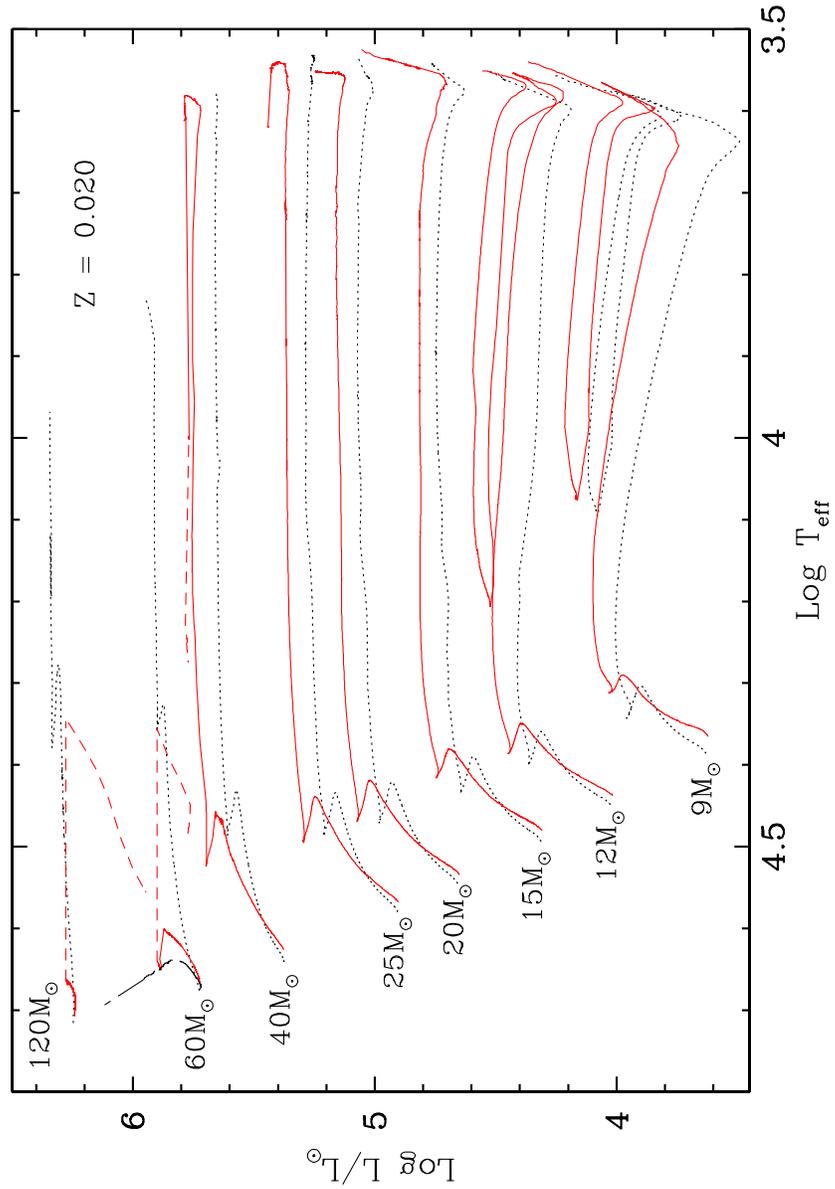}
\caption{Evolutionary tracks for non-rotating 
(dotted lines) and
rotating (continuous lines)  models with solar metallicity
(Meynet \& Maeder 1999). 
The rotating models
have an initial velocity $v_{\rm rot}$ of 300 km/s.
The dashed track corresponds to a very fast rotating star
($v_{\rm rot} \sim 400$ km/s) of 60 M$_\odot$, which
follows a nearly homogeneous evolution.}
\end{figure}

\begin{figure}
\epsfxsize=12cm  \epsfbox{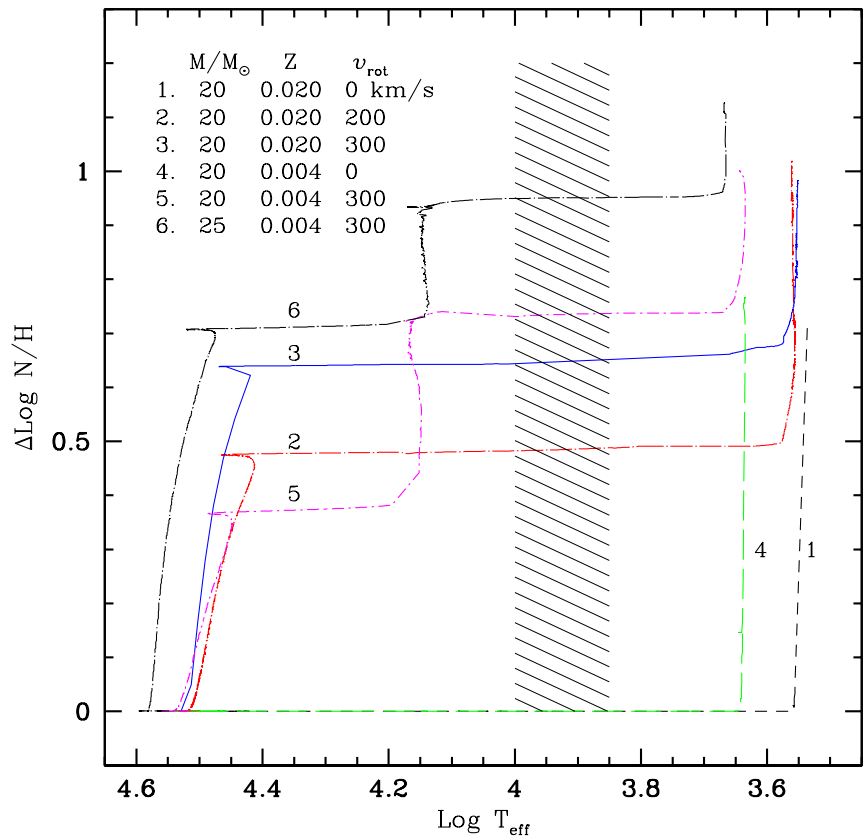}
\caption{Evolution as a function of log $T_{\rm eff}$ of
$\Delta \log {N\over H}= \log (N/H)-\log (N/H)_i$, where $N$ and $H$ 
are the surface abundances (in number) of nitrogen and hydrogen 
respectively, the index $i$ indicates initial values. 
The initial masses, metallicities and
rotational velocities are indicated. The shaded area corresponds to the range
of observed values (Venn 1999) for A-type supergiants in the SMC .}
\end{figure}

\begin{figure}
\epsfxsize=12cm  \epsfbox{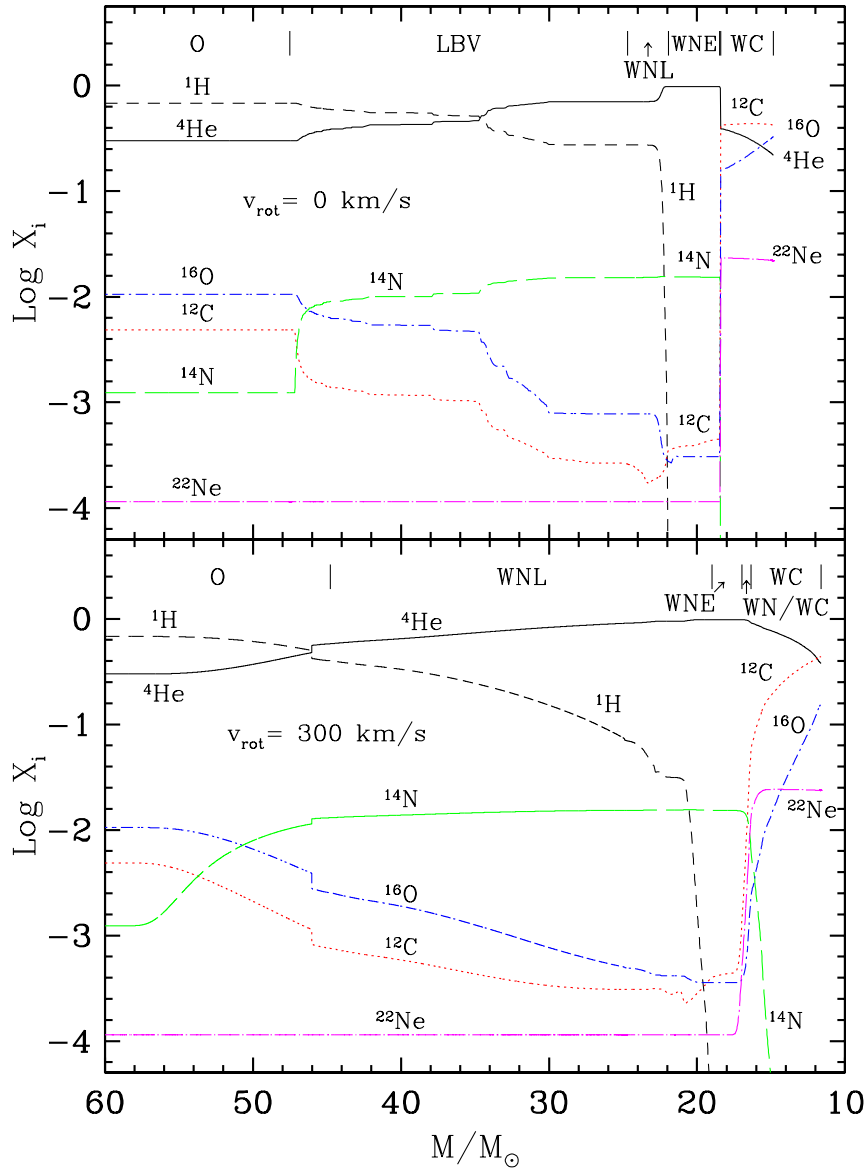}
\caption{Evolution of the abundances at the surface of a
60 M$_\odot$ star as a function of the remaining stellar mass
for different initial rotational velocities $v_{\rm rot}$. 
The parts
of the evolution during which the star may be considered
as an O-type star, a LBV and a W--R star are indicated. During
the W--R phase, the WN, the transition ``WN/WC'' and 
the WC phases are distinguished.}
\end{figure}

\end{document}